\documentclass{emulateapj}
\usepackage{graphicx}
\usepackage{xspace}
\usepackage{natbib}
\usepackage{color}
\usepackage{url}

\usepackage{setspace}

\newcommand{\g}{$\gamma$}
\def\arcsec{$^{\prime\prime}$}

\def\flux{erg cm$^{-2}$ s$^{-1}$}

\renewcommand{\deg}{\ensuremath{^{\circ}}\xspace}
\begin{document}

\title{ Hunting for treasures among the {\it Fermi} unassociated sources: a multi-wavelength approach}
\shorttitle{Treasures among {\it Fermi} Unassociated Sources}

%\author[1]{Fabio Acero}
%\author[1]{Davide Donato}
%\author[1]{Roopesh Ojha\altaffilmark{1,2}}
%
%\affil[1]{ORAU/NASA Goddard Space Flight Center, Astrophysics Science Division, Code 661, Greenbelt, MD 20771}
%

\author{F. Acero\altaffilmark{1}, D. Donato\altaffilmark{2}, R. Ojha\altaffilmark{1,3,4}, 
J. Stevens\altaffilmark{5}, P. G. Edwards\altaffilmark{6}, E. Ferrara\altaffilmark{2}, J. Blanchard\altaffilmark{7}, J. E. J. Lovell\altaffilmark{7}, D. J. Thompson\altaffilmark{8}  }

\altaffiltext{1}{ORAU/NASA Goddard Space Flight Center, Astrophysics Science Division, Code 661, Greenbelt, MD 20771}
\altaffiltext{2}{CRESST/NASA Goddard Space Flight Center, Astrophysics Science Division, Code 661, Greenbelt, MD 20771}

\altaffiltext{3}{Adjunct Professor, The Catholic University of America, 620 Michigan Ave., N.E.  Washington, DC 20064}
\altaffiltext{4}{Honorary Fellow, University of Tasmania School of Mathematics \& Physics Private Bag 37 Hobart TAS 7001, Australia}

\altaffiltext{5}{CSIRO Astronomy and Space Science, Locked Bag 194, Narrabri NSW 2390, Australia}
\altaffiltext{6}{CSIRO Astronomy and Space Science, PO Box 76, Epping NSW 1710, Australia}
\altaffiltext{7}{University of Tasmania School of Mathematics \& Physics Private Bag 37 Hobart TAS 7001, Australia}
\altaffiltext{8}{NASA Goddard Space Flight Center, Astrophysics Science Division, Code 661, Greenbelt, MD 20771}

%
%\author{Davide Donato}
%\altaffiltext{2}{CRESST/NASA Goddard Space Flight Center, Astrophysics Science Division, Code 661, Greenbelt, MD 20771}
%
%\author{Roopesh Ojha\altaffilmark{1,2}}
%\affil{ORAU/NASA Goddard Space Flight Center, Astrophysics Science Division, Code 661, Greenbelt, MD 20771}
%
%\author{Jamie Stevens}
%\affil{CSIRO Astronomy and Space Science, Locked Bag 194, Narrabri NSW 2390, Australia}
%
%\author{Philip G. Edwards}
%\affil{CSIRO Astronomy and Space Science, PO Box 76, Epping NSW 1710, Australia}
%
%\author{Jay Blanchard and James E. J. Lovell}
%\affil{University of Tasmania, Newnham Dr, Newnham TAS 7248, Australia}
%
%\author{Elizabeth Ferrara}
%\affil{CRESST/NASA Goddard Space Flight Center, Astrophysics Science Division, Code 661, Greenbelt, MD 20771}
%
%\author{David Thompson}
%\affil{NASA Goddard Space Flight Center, Astrophysics Science Division, Code 661, Greenbelt, MD 20771}
%

\shortauthors{Acero et al.}
%\affil{NASA/GSFC}
\email{fabio.f.acero@nasa.gov}

%%\author{D. Donato\altaffilmark{1,2} and B. Cenko\altaffilmark{3}}
%%\email{davide.donato-1@nasa.gov}
%%
%%\author{S. Covino\altaffilmark{4}, E. Troja\altaffilmark{1}, T. Pursimo\altaffilmark{5}, C. C. Cheung\altaffilmark{6}, O. Fox\altaffilmark{3}, A. Kutyrev\altaffilmark{7}, S. Campana\altaffilmark{4}, D. Fugazza\altaffilmark{4}, H. Landt\altaffilmark{8}}
%%
%%\and
%%
%%\author{The RATIR Team\altaffilmark{9}}
%%
%%
%%%% Specify alternate affiliation information with \altaffiltext, 
%%%% with one command per each affiliation.
%%
%%\altaffiltext{1}{CRESST and Astroparticle Physics Laboratory NASA/GSFC, Greenbelt, MD 20771, USA.}

%%%%%%%%%%%%%%%%%%%%%%%%%%%%%%%%%%%%%%%%%%%%%%%%%%
\begin{abstract}

%The {\it Fermi} Gamma-ray Space Telescope has been providing a wealth of high-interest sources where the multi-wavelength counterpart was either inconclusive or missing altogether. We report on a number of new detections of active galactic nuclei resulting from follow-up observations of high-galactic latitude {\it Fermi} targets in the southern hemisphere. We selected these targets because their multi-wavelength properties made them strong candidates to contain AGN. We discuss the selection criteria for the sources, their associations with new AGN, and the success rate of our approach.

The {\it Fermi Gamma-ray Space Telescope} has been detecting a wealth of
 sources where the multi-wavelength counterpart is
either inconclusive or missing altogether. We present a combination of
factors that can be used to  identify multi-wavelength counterparts to these {\it Fermi} unassociated sources. This approach was
used to select and investigate seven bright, high-latitude unassociated sources with radio, UV, X-ray and $\gamma$-ray observations.
As a result, four of these sources are candidates to be active galactic nuclei (AGN), and one 
to be a pulsar, while two do not fit easily into these known categories of sources. 
The latter pair of extra-ordinary sources might reveal a new category subclass or a new type of $\gamma$-ray emitters.
These results altogether demonstrate the power of a multi-wavelength approach to illuminate the nature 
of  unassociated {\it Fermi} sources. 

%Unique counterparts were determined for four of them, while two others have
%multiple possible counterparts. 

%The X-ray and radio limits on the unassociated Fermi, with no obvious counterpart, hint at a possibly
%exotic nature for this object.

\end{abstract}

%%%%%%%%%%%%%%%%%%%%%%%%%%%%%%%%%%%%%%%%%%%%%%%%%%
\section{Introduction}

The ongoing survey of the \g-ray sky with the {\it Fermi} Large Area Telescope (LAT) has 
led to a tremendous increase in the known population of \g-ray sources. The 2-year 
{\it Fermi}-LAT catalog \cite[2FGL;][]{nolan12} lists 1873 sources, divided into the following classes: 1) active galactic nuclei (AGN ; $\sim$60\%);
2) pulsars and binary systems (6\%); 3) supernova remnants, pulsar wind nebulae and 
other Galactic sources (4\%);  and unassociated sources ($\sim$30\%). Thus the 
second largest `class' consists of sources that do not have a clear association with 
plausible counterparts at other wavelengths. At the very least, this large number of 
mysterious \g-ray sources includes outliers of known classes of sources. It is not 
unreasonable to suppose that a subset of these sources could provide a pathway to new discoveries.
% and the \g-ray sky as seen by the {\it Fermi}-LAT still holds mysterious treasures waiting to be unveiled. 

The unassociated sources in the 2FGL catalog at high Galactic latitude 
($|b|>5\degr$) are likely to fall under two main categories: AGNs or millisecond pulsars (MSPs).
 The members of each category share characteristic \g-ray properties that can be used as 
patterns to evaluate the probability of a given source to belong to a class of 
objects \cite[e.g.,][]{ackermann12unassoc,mirabal12}. The typical \g-ray 
signature for an AGN is non-periodic variability and a power-law or broken 
power-law spectral shape \citep[see, e.g., the second {\it Fermi-LAT} catalog of AGN, ][]{ackermann11-2lac}.
For pulsars and MSPs, the key to identification is the detection of pulsations in \g-rays. Further, their 
spectra are usually curved and present a cut-off around a few GeV  
\citep[see, e.g., the first {\it Fermi} pulsar catalog,][]{abdo10-1pc}.

Although the source localization accuracy made possible by the {\it Fermi-LAT}
 has greatly improved over previous \g-ray instruments, the typical 95\% 
confidence level position uncertainty radius (R$_{95}$) of 2FGL sources is of the order of
$\sim0.1\degr$, still relatively large  with respect to lower energy positional errors
 to make associations solely based on the \g-ray position.

To resolve this issue, multi-wavelength campaigns can be decisive in revealing 
the nature of some {\it Fermi} unassociated sources.
%have often been the trigger/key element 
%Multi-wavelength follow-up campaigns have proven to be a key element to unveil the nature of some {\it Fermi} unassociated sources.
For example, the deep search for radio pulsations at \g-ray positions has led to
the discovery of unexpectedly large numbers of pulsars and MSPs \citep{ray12}
 in the population of \g-ray unassociated sources.\\

In addition, X-rays are very good tracers of energetic processes, and above 
a few keV they are not affected by absorption along the line of sight. From
the observational point of view, the current X-ray satellites have 
 relatively large fields of view (FoV), high sensitivity and a 
localization at the arcsec level. 

A precise location, derived from the X-rays, can also strongly enhance the sensitivity of \g-ray blind 
searches for pulsation of isolated pulsars \citep{dormody11} with {\it Fermi}-LAT by dramatically reducing the numbers of 
position trials. In the case of an MSP in a binary system, optical observations of the companion are needed to 
obtain the orbital period of the system \citep[see][for a recent example]{romani12}.
However, it is often found that there is more than one plausible
X-ray counterpart to a {\it Fermi} unassociated source.
{ While this number can largely vary depending on the position on the sky and the characteristics of the telescope, \citet{cheung12}, for example, found $\sim$10 sources with the \textit{Chandra} X-ray telescope in the error ellipses of the two brightest 0FGL unassociated sources  0FGL J\,1311.9-3419 and 
0FGL J\,1653.4-0200. }
In those cases, the high spatial resolution and large frequency range provided by radio interferometry can  provide a crucial step in
identifying the counterpart. Observations at multiple radio frequencies are useful as they can indicate the nature
of the object identified, e.g., relatively high compact radio flux density and a flat radio spectrum suggest the object could be an AGN.
Once a potential AGN counterpart has been identified, optical follow-up to search for a redshift and optical polarization 
will provide another strong piece of evidence. Unambiguous identification is made when correlated variability is observed in $\gamma$ rays and another wavelength.

The identification of {\it Fermi} unassociated sources with known classes
of astrophysical objects is of interest, not least because these new
identifications often constitute a new population, e.g. AGN
identifications might discover a subclass of radio-weak, $\gamma$-ray loud objects.

{
On a population scale, several groups \citep[see e.g.][]{ackermann12unassoc,mirabal12} have developed methods to try to predict the nature of a {\it Fermi} 
unassociated source based on a subset of its $\gamma$-ray 
features that were found to be effective at discriminating between AGN and pulsars. 
\citet{mirabal12}, for example, used the spectral and variability 
information about the LAT sources to assign a probability P(AGN) that an 
unassociated source is an AGN.
Other approaches have studied the  lower wavelength counterparts to the $\gamma$-ray sources.
For example, \citet{massaro12a,massaro12b} have developed a method to identify the AGN candidates among the
2FGL unassociated sources based on the colors of the infrared counterparts lying within the {\it Fermi} error ellipse using 
 the all-sky survey from the \textit{Wide-field Infrared Survey Explorer} (\textit{WISE}). }
 
%Using the all-sky survey from the \textit{Wide-field Infrared Survey Explorer} (\textit{WISE}), \citet{massaro12a,massaro12b}
%have developed a method to identify the AGN candidates among 
%2FGL unassociated sources based on the colors of the infrared counterparts lying within the {\it Fermi} error ellipse.   }

{While the largest class of predicted sources from all these methods is AGN, }
some sources did not fit in the AGN or pulsars category, and were considered ``unclassifiable'' or ``outlier objects''. 
It is vital to make these identifications so that the properties of these
truly ``exotic" {\it Fermi} unassociated sources can be studied.\\

%Our study presents the results from new radio, optical, UV and X-ray observations and ????

%%%%%%

%However,  in the methods to predict the nature of unassociated sources discussed above
% some sources did not fit in the AGN of pulsars category, and were considered 
%``unclassifiable'' or ``outlier objects''. 

%Their Random Forest classifier, $\text{SIBYL}$, predicts 216 objects to be AGN,
%16 to be pulsars and 37 without a firm prediction \citep{mirabal12}. Where available, we
%have listed these predicted probabilities to be an AGN in Table \ref{table}.
%

 \medskip

%%%%%%%%%%%%%%%%%%%%%%%%%%%%%%%%%%%%%%%%%%%%%%%%%%%%%%%%%%%%%%%%%%%%%%%%%%%%%%%
\section{Description of the method}
\label{selection}

\subsection{Selection of {\it Fermi}-LAT sources}

To limit the potential multi-wavelength counterparts to explore, we selected  bright
unassociated  \g-ray sources (average significance in all 
energy bands given in the 2FGL $\sigma >$10), as these usually have smaller position 
uncertainty.  We selected sources located at high latitude ($|b|$$>$5$\degr$) where 
the density of potential counterparts is lower. We removed all unassociated sources 
labeled as ``c" in 2FGL (those found in a region with bright and/or possibly incorrectly modeled diffuse 
emission).  Those different selection criteria reduced the sample of unassociated
{\it Fermi} sources in the 2FGL from 555  to 85 ($ \sigma >$10), then 33 ($|b|$$>$5$\degr$) and finally 32 (no ``c" flag). 

Amongst this sample of bright, high latitude and unconfused unassociated 2FGL targets, we 
selected the sources that had been observed in the X-rays by the {\it Swift} satellite, whose X-ray telescope (XRT)
FoV typically encompasses the {\it Fermi} uncertainty ellipse. We excluded 
sources observed by other recent X-ray missions (i.e., {\it XMM}, {\it Suzaku}, 
and {\it Chandra}) because these data sets have already been  studied extensively.
{In 2010, {\it Swift}  began a systematic search\footnote{http://www.swift.psu.edu/unassociated} for X-ray counterparts of {\it Fermi}-LAT unassociated sources 
\citep[see][]{stroh13}.} For this reason, the large majority of those bright, high latitude unassociated 2FGL targets have
X-ray coverage. We selected only observations with at least 3.5 ksec  exposure time. Our final list includes 22 $\gamma$-sources.

{\subsection{Identification of X-ray counterparts}}

{The \textit{Swift} observations were analyzed (see Sect. \ref{xray}) to identify potential X-ray counterparts
within the {\it Fermi} R$_{95}$ error ellipse. Among the 22 \g-ray sources, 16 had at least one  X--ray counterpart 
detected with a significance greater than 3$\sigma$. 
The remaining 6 \g-ray sources\footnote{2FGL J\,0032.7$-$5521, J\,0934.0$-$6231, J\,1625.2$-$0020,  J\,1744.1$-$7620, J\,2039.8$-$5620, and J\,2112.5$-$3042.} had no X-ray counterpart and represent a puzzling sample that would require deeper X-ray observations.
 The spatial resolution of the {\it Swift} XRT (18'' half power diameter)  allows a localization at the level of a few arcseconds. 
 Once this precise position  in the X-ray is obtained, multi-wavelength follow-ups with radio, IR, or optical observatories
 (whose FoV is typically smaller) are possible. }

\subsection{Radio follow-up of potential counterparts}

The combination of X-ray and follow-up radio observations provides a powerful 
tool to test the two most common known scenarios (AGN or pulsar and MSP) for our sample of 
{\it Fermi} unassociated sources. Sources that do not show typical characteristics of these two populations 
are therefore promising candidates for new discoveries.

Multi-frequency radio observations of X-ray counterparts are useful to search 
for radio flat spectrum sources, a typical signature for an AGN. If no radio 
counterpart to the X-ray source is detected (or is faintly detected only at the lowest radio 
frequency), the \g-ray source might be a good MSP candidate.
Obtaining the precise X-ray positions of the potential counterparts was a key 
element to optimize our radio observation strategy.  These X-ray positions
were used to carry out follow-up observations with the Australia Telescope Compact Array (ATCA) radio telescope and 
to minimize the number of radio pointings per {\it Fermi} source. This approach
allowed deeper observation of each X-ray counterpart. The typical {\it Fermi} 
error ellipse is 0.1$\degr$ in the semi major axis and the primary beam of the 
ATCA radio telescope at the observed frequencies ranges from about 10 arcmin 
at 5.5\,GHz to about 1 arcmin  at 40\,GHz.

From the list of 16 \g-ray sources with at least one X-ray counterpart, radio observations of 7 sources were
 obtained with ATCA. These objects are presented  in Table \ref{table}.

\begin{table*}[t]
\label{table}

\caption{ }

\setlength{\tabcolsep}{1.8pt}
\small
\begin{center}
\begin{tabular}{lccc|c|c|cc|cc|ccc|c}
%\begin{tabular}{lccccccccccccc}

\hline

%\\[1cm]

\vspace{0.15cm}
2FGL\, name  &  \multicolumn{3}{c}{J\,0143.6} &   \multicolumn{1}{c}{J\,0523.3} &   \multicolumn{1}{c}{J\,0803.2} &  \multicolumn{2}{c}{J\,1036.1} &  \multicolumn{2}{c}{J\,1129.5} &  \multicolumn{3}{c}{J\,1231.3} & J\,1844.3  \\[-0.05cm]
\hline %\\[-0.15cm]

P(AGN) & \multicolumn{3}{c|}{1.0}     &  \multicolumn{1}{c|}{0.738}  &   \multicolumn{1}{c|}{0.964} &  \multicolumn{2}{c|}{-} &  \multicolumn{2}{c|}{0.962} &  \multicolumn{3}{c|}{0.554} & 0.996  \\[0.15cm]

X-ray name    &   A1 &   A2 &   A3 &   B1 &   C1 &   D1 &   D2 &   E1 &   E2 &   F1 &   F2 &   F3  & G1    \\

~~~~~~~~ $\sigma_{\rm det}$ & 43.0 &  3.3 &  3.0 &  3.3 &  9.6 &  5.6 &  3.1 &  3.4 &  3.0 & 3.9 &  3.3 &  3.6 & 16.8   \\
%\vspace{0.15cm}

~~~~~~~~ $\sigma_{\rm var}$ &  4.8 &    N &    N &    N &  3.5 &    N &    N &    N &    N &  - &  - &  -  & 9.1  \\[0.15cm]

UV    & 1- -1 & 1- -1 & 1- -1 & -0- - & -1- - & 1100   & 1111       & -111      & -111      &  1111        & 1-1-       & 1-11  & - -1-      \\[0.15cm]
%IR & & & & & & & & & & & & & \\

%\vspace{0.15cm}
radio & 111100 & 100000 & 100000 & 100000  & 111110  & 10- - - - & 00- - - - & 11- - - - & 00- - - - &  00- - - -  & 00- - - - & 00- - - -  & 111111  \\

%FOM & 0.90? & & & & & & & & & & & & \\

\hline
\end{tabular}
\end{center}

\scriptsize{Multi-wavelength properties of the observed sample of 2FGL 
unassociated sources. For each \g-ray source we report the probability of the 
source to be an AGN based on its GeV properties \citep{mirabal12}, see Sect. \ref{discussion} for more details. For each
X-ray source within or very close to the \g-ray uncertainty ellipse, we report 
1) the label as shown in Fig. \ref{image}; 2) the detection significance 
($\sigma_{\rm det}$) of the
X-ray source; 3) the significance of X-ray variability. When no significant 
variability was observed ($\sigma_{\rm var}<$3) or the variability could not be 
tested (only one observation available), we used the label \textit{N} and 
a dash, respectively; 4) the detection of an 
UV counterpart in the U, W1, M2, and W2 filters; 5) the detection of a radio 
counterpart at 5.5, 9, 17, 19, 38, and 40 GHz. For both the UV and radio 
rows, ``1'' and ``0'' represent a detection and an upper-limit, respectively. A
dash means that the source was not observed in that specific filter/frequency.
The sources A3, B1 and D1  have a faint radio counterpart just above the detection
threshold and although they are not detected at high frequency a radio flat 
spectrum is not excluded. The source J\,1129.5$+$3758 could not be observed at higher radio 
frequencies due to its low elevation angle.} 

\label{table}
\end{table*}

%%%%%%%%%%%
\section{Data analysis}

%%%%%%%%%%%
%\subsection{Swift}

\subsection{X-rays}
\label{xray}

We analyzed all the archival clean event files obtained in the photon counting
mode by the XRT onboard {\it Swift} and covering the uncertainty ellipse of each
\g-ray source. In some cases several exposures of the same \g-ray source have 
been taken. After combining them within \verb+XSelect+, we used 
\verb+Ximage+ to perform source detection.

% The typical X-ray uncertainty  position is of the order of 3\arcsec-6\arcsec.

We then used those coordinates to run ``The {\it Swift}-XRT data products 
generator'' available at the University of Leicester 
website\footnote{\url{http://www.swift.ac.uk/user$\_$objects/index.php}} to perform the data
reduction and the early data analysis. This facility allowed the creation of 1) 
three combined event files of all the observations, one in the full energy band 
0.3--10 keV and two in the sub-bands 0.3--1.5 keV (S band) and 1.5--10 keV (H band) ; 2) a combined
0.3--10 keV spectral file for both the source and the background; 3) a 0.3--10 
keV light curve, binned by observation.  We used the 0.3--10 keV combined event 
file to estimate the significance of detection ($\sigma_{\rm det}$), the combined
spectral file to perform a spectral analysis for those sources with a high
level of detection significance, the sub-bands event files to estimate the 
hardness ratios (defined as $\frac{H-S}{H+S}$, where the counts are typically
extracted from a 20\arcsec circular region) for the less significant sources, and
the light curve to check the presence of variability. The wide range of separation 
between {\it Swift} snapshots (from a few days to many months) has allowed us to 
test the X-ray variability of potential, bright counterparts, by comparing the X-ray flux in 
the 0.3--10 keV energy band for each available observation. 
The significance of the X-ray variability was calculated as 
$\sigma_{\rm var}= (CR_{\rm max}  - CR_{\rm min })/\sqrt{CR_{\rm min,err}^{2}+CR_{\rm max,err}^{2}  } $  
where $CR$ is the net count rate.  
{We note that  all the detected sources are within 10 arcmin of the center of the images, where the XRT is best calibrated,
and the estimated count rates are not affected by issues at the edges of the CCD.
The values of $\sigma_{\rm var}$ and $\sigma_{\rm det}$ are reported in Table \ref{table} 
for X-ray sources found  within the R$_{95}$ {\it Fermi} error ellipse at a detection significance $> 3 \sigma$. 
The position of the X-ray sources presented in Sect. \ref{xray-results} were obtained with the task  {XRTcentroid}.}
%\Delta (CR) / \sigma_{\rm tot} =

%($\sigma_{\rm var}$) is reported in Table \ref{table}. 
%sigma_var = Delta (c/r) / sigma_tot  = (c/r_max - c/r_min ) / sqrt( c/r_err_min^2 + c/r_err_max^2)
%where CR = net count rate.

\subsection{UV and optical}
\label{UV}

The Ultraviolet/Optical Telescope (UVOT)
provides coverage simultaneous to the XRT using the UVOT ``filter of the 
day''. This gives us a partial, random coverage in the 
U, W1, M2, and W2 filters (no observations with V or B filters were performed). We analyzed the public data using the 
standard tools from the \textit{Swift} analysis Web page of HEASARC\footnote{http://heasarc.nasa.gov/docs/swift/analysis/}. 

For each filter, we combined all the exposures within a single 
observation to estimate the monochromatic flux (corrected for the 
finite aperture of the extraction region, coincidence loss, and large 
scale sensitivity).
For sources with multiple observations we kept them separated to check for variability in the bright sources.
For the dim sources, the observations were combined  to increase the signal-to-noise ratio or to
lower the upper limit in case of non-detection. We summed both the sky images 
and the exposure maps using \verb+uvotimsum+. The photometry has been obtained
by running \verb+uvotsource+ and using a circular source extraction region 
(with radius varying between 3\arcsec and 5\arcsec, depending on the source 
intensity) and an annular background centered on the source (with an inner 
radius not less than 15\arcsec). In Table \ref{table} we report if the source 
has been detected in each available filter. The detection threshold in 
\verb+uvotsource+ has been set to the canonical 3$\sigma$.

%%%%%%%%%%%
\subsection{Radio}
\label{radio}

Regular observations with the ATCA
are a key component of the TANAMI program \citep{Ojha10}, which monitors 
southern hemisphere {\it Fermi}-LAT detected sources at a
number of radio frequencies and resolutions. Every few weeks,
``snapshot" observations are made at the frequencies 5.5, 9, 17, 19,
38 and 40\,GHz, where each frequency is the center of a 2\,GHz wide
band and the fluxes are calibrated against the ATCA primary flux
calibrator PKS\,1934$-$638 \citep{stevens12}.
 
For some candidates there were multiple X-ray counterparts within the {\it Fermi}-LAT
error circle, and in many cases these counterparts were too far from
each other to be observed in one pointing of ATCA. Each observation was for 8
minutes at each band with the first observation made at the lowest
frequency, where an AGN counterpart would likely have the highest
flux density. In cases where no detection was made at the lower frequencies no observations were made
at the higher frequencies.

These positions and error ellipses were obtained by first imaging each
source (that had at least some detected flux at 5.5\,GHz) in a
standard manner using MIRIAD. Due to the elongated beams (in most
cases) the beam fitting was done using the task IMFIT. IMFIT was used
to obtain the position and error ellipse after checking that the
source was realistically extracted from the image.

\section{Results}

The results of our X-ray and radio observational campaigns are described in the following sections.
These results are  summarized in Table \ref{table}, and the UV, radio and X-ray fluxes are provided in an online electronic table.

\subsection{Swift results}
\label{xray-results}

While some of the \g-ray sources have only one
detected counterpart, for the majority of them multiple X--ray counterparts have
been found within, or just outside, the R$_{95}$ \g-ray position uncertainty 
reported in the 2FGL catalog.
Only a handful of them have a detection significance above 5$\sigma$.
The typical X-ray uncertainty  position is of the order of 3\arcsec-6\arcsec.

Here we report the results of the analysis of the XRT and UVOT instruments.
Whenever the statistics were sufficient, a  double-absorbed power-law model was
fitted to the data to derive basic spectral parameters for the sources detected
with XRT. The first absorption corresponds to the Galactic value \citep[from 
the Leiden/Argentine/Bonn Survey of Galactic HI;][]{kalberla05} and the second 
is intrinsic to the source and is left free to vary in the fitting. 
{Both absorptions were modeled using the \textit{phabs} model available in XSPEC.}

\medskip

%\begin{itemize}

%\item
 {\bf J\,0143.6$-$5844}: In the R$_{95}$ {\it Fermi}-LAT ellipse we found 3 X--ray
sources in a 4.4-ksec exposure. While A2 and A3 (RA$=01^h 43^m 49.85^s$; 
Dec=$-58\deg 43' 19.2''$; and RA=$01^h 43^m 38.61^s$; Dec=$-58\deg 41' 
50.2''$) have a low detection 
significance and are moderately soft (hardness ratio HR=-0.6 and -0.3, respectively), A1 
(RA$=01^h 43^m 47.57^s$; Dec=$-58\deg 45' 51.6''$) is a bright object with 
1590 counts. The spectral analysis revealed an  absorption along the line of sight in slight excess over the Galactic value (n$_{H,Gal}=2.04\times
10^{20}$ cm$^{-2}$, n$_{H,int}=3.3_{-0.2}^{+6.5} \times 10^{20}$ cm$^{-2}$) with a photon index $\Gamma=2.29\pm0.12$. \textit{Swift} observed the
field of view of this \g-ray source 3 times. While the hardness ratio of A1 does 
not change over time (indicating no spectral evolution), the net, corrected 
count-rate varies between 0.503$\pm$0.024 and 0.671$\pm$0.026, with a 
significance of variability of $\sigma_{var}\sim4.8$. Both the brightness 
changes and the shape of the spectrum are typical characteristics of an AGN. 
All 3 sources were detected at UV frequencies in the U and W2 filters.
In the first and last observation, both performed with the W2 filter, no sign 
of flux variation is seen for A1 {while only one observation was performed with the U filter and no variability can be assessed.}

\medskip

%\item
 {\bf J\,0523.3$-$2530}: The only possible X--ray counterpart detected by 
\textit{Swift} in a 4.8-ksec exposure is a dim object (RA$=05^h 23^m 17.11^s$; 
Dec=$-25\deg 27' 31.9''$), not found in the UV band. 

\medskip

%\item 
{\bf J\,0803.2$-$0339}: Also in this case, only one X--ray source has been
found (RA$=08^h 03^m 12.11^s$; Dec=$-03\deg 36' 1.4''$) but at relatively 
high significance. The 80 counts detected in 3.9 ksec were distributed following
a power law of spectral index $\Gamma= 2.10_{-0.31}^{+0.60}$. 
\textit{Swift} performed two observations on this object, and the
light curve analysis shows some sign of variability ($\sigma_{var} = 3.5$) with
the count-rate changing from 0.018$\pm$0.003 to 0.039$\pm$0.005. Also in this
case, no significant optical variability has been found in the two observations
with the W1 filter.

\medskip

%\item
 {\bf J\,1036.1$-$6722}: The field of view of this source has been observed 
extensively (35.5\,ksec, spanning 2 years) but no bright X--ray sources have 
been detected. Only two marginal sources can be found above the 3-$\sigma$
threshold: D1 (RA$=10^h 35^m 45.98^s$; Dec=$-67\deg 25' 15.4''$) and D2 
(RA$=10^h 36^m 22.13^s$; Dec=$-67\deg 22' 28.5''$). The sources are very faint
also in the UV band and are detected only at longer wavelength thanks to the 
long exposures. Due to the low statistics, no variability analysis can be performed. The number of X--ray
counts for the two sources is 34 and 10, respectively. But while in D1 they are
evenly split below and above 1.5 keV ($\Gamma=1.9_{-0.8}^{+0.7}$), the counts are
only in the soft band in D2. 

\medskip

%\item
 {\bf J\,1129.5$+$3758}: In 4 pointings (for a total of 4.8\,ksec) only
two faint X--ray sources were detected: E1 (RA$=11^h 29^m 03.42^s$; 
Dec=$+37\deg 56' 58.6''$) and E2 (RA$=11^h 29^m 31.12^s$; Dec=$+38\deg 01' 
59.6''$). The total counts are 11 and 10, distributed in a 2:1 ratio between
the soft and hard band (HR =$-0.3$ and $-0.4$). Both sources are detected in the UV
range.

\medskip

%\item
 {\bf J\,1231.3$-$5112}: Also in this case only dim objects were detected.
F1 (RA$=12^h 31^m 51.31^s$; Dec=$-51\deg 19' 39.9''$), F2 (RA$=12^h 30^m 52.13^s$; Dec=$-51\deg 19' 17.1''$),
and F3 (RA$=12^h 31^m 29.62^s$; Dec=$-51\deg 09' 32.3''$) 
have 16, 5, and 15 counts observed in a 7.3\,ksec exposure. F1 is a hard source
(HR = $0.25$), while the other two sources have counts in the soft band only. While
F1 and F2 have dim optical counterparts, F3 is a bright star, making this source
an unlikely association of the \g-ray emission.

\medskip

%\item
 {\bf J\,1844.3$+$1548}: Both 
the X--ray and UV analyses suggest that the likely counterpart of the \g-ray 
source is an AGN. The X--ray position (RA$=18^h 44^m 25.42^s$; Dec=$+15\deg 46'
44.3''$) coincides with a known radio source (NVSS J\,184425$+$154646). The 
spectral analysis of the 4.2\,ksec exposure indicates a very steep photon index 
($\Gamma = 3.00_{-0.39}^{+0.42}$), an intrinsic absorption well above the Galactic
value (n$_{H,Gal} = 1.73 \times 10^{21}$, and n$_{H,int} = 2.63\pm0.06 \times 
10^{21}$), and a 0.3--10 keV unabsorbed flux F$_X = 9.5_{-3.2}^{+6.7} \times 
10^{-12}$ \flux. The X--ray evolution indicates that the source count-rate 
varied between 0.110$\pm0.007$ on 2011 Nov. 18 and 0.019$\pm$0.007 on 2011 Nov. 
30, and it remained constant at (0.020$\pm$0.006) on 2011 Dec. 4. The UV light
curve shows variation as well: in the M2 filters the source brightness declined 
by a magnitude from the first (m$_{M2}=19.00\pm0.08$) to the last observation
(m$_{M2}=19.99\pm0.25$).

%\end{itemize}

\subsection{Radio results}

Seven of our 16 {\it Fermi} unassociated sources were observed over two epochs on 27
August 2012 and 4 September 2012. Altogether, a total of 13
radio pointings were required to observe all the possible X--ray counterparts. 

Six of our seven {\it Fermi} sources were detected in at least one radio
frequency at one of their X-ray candidate counterpart locations.
The flux density of the sources detected ranges from 8 mJy to 97 mJy (source B1 and F1 respectively) at 5.5 GHz.
All the radio sources reported in Table \ref{table} are within the 1$\sigma$ error ellipse of their 
corresponding X-ray source.
Unique radio counterparts to the X-ray detection were found for the four sources 
(A1, C1, E1, G1) that  were detected in multiple radio bands. 

{In the case of the sources J\,0523.3$-$2530 and J\,1036.1$-$6722 the error ellipse of our radio observation includes three discrete sources identified by earlier radio surveys. Thus we cannot determine which of these three sources (or which two or all three) are being detected.
Higher resolution observations and/or multi-epoch variability observations will be required to firmly establish the counterpart to these two {\it Fermi} candidates.}

For  J\,0143.6$-$5844 in addition to A1, two alternate nearby radio sources (A2 and A3) were detected 
with a much lower radio flux density,  almost certainly due to the large beam\footnote{While pointed at A2 and A3, the telescope is sensitive to flux from A1.} at 5.5\,GHz. 
J\,1129.5$+$3758 was observed to have a flat spectrum between two observed frequencies (5.5 and 9.0\,GHz). Observations at higher frequencies were not performed due to its low elevation at ATCA, which would make such observations problematic. We will confirm that its flat spectrum extends to higher frequencies in future observations with a more northern telescope. 
The only \g-ray source where no radio sources were detected is J\,1231.3$-$5112 
and an upper limit of $\sim$ 10 mJy at 5.5 and 9\,GHz can be placed on all three radio pointings.

In follow-up observations on 26 August 2012 with the two element Ceduna Hobart Interferometer \citep[CHI, ][]{blanchard12} which has a resolution of 6.6 milliarcsecond at the observed frequency of 6.7\,GHz, source A1 in J\,0143.6-5844 was detected with a flux density of $25\pm4$ mJy. This compares to a flux density of $26\pm3$ mJy measured by ATCA at the adjacent frequency of 5.5\,GHz. Thus, we conclude that J\,0143.6-5844 is a very compact object.  CHI observations of the other detected sources are also planned.

An independent radio follow-up of unassociated 2FGL sources by \citet{petrov13}, found the same candidate counterparts for J\,0143.6-5844, 
J\,0523.3$-$2530 and J\,0803.2$-$0339 and confirmed our non detection of J\,1231.3$-$5112.

\section{Discussion}
\label{discussion}

\subsection{AGN candidates}

\begin{figure*}
 \centering

   \begin{tabular}{cc}

%\hspace{-0.5cm}
\includegraphics[width=6cm,bb=23 188 558 680,clip]{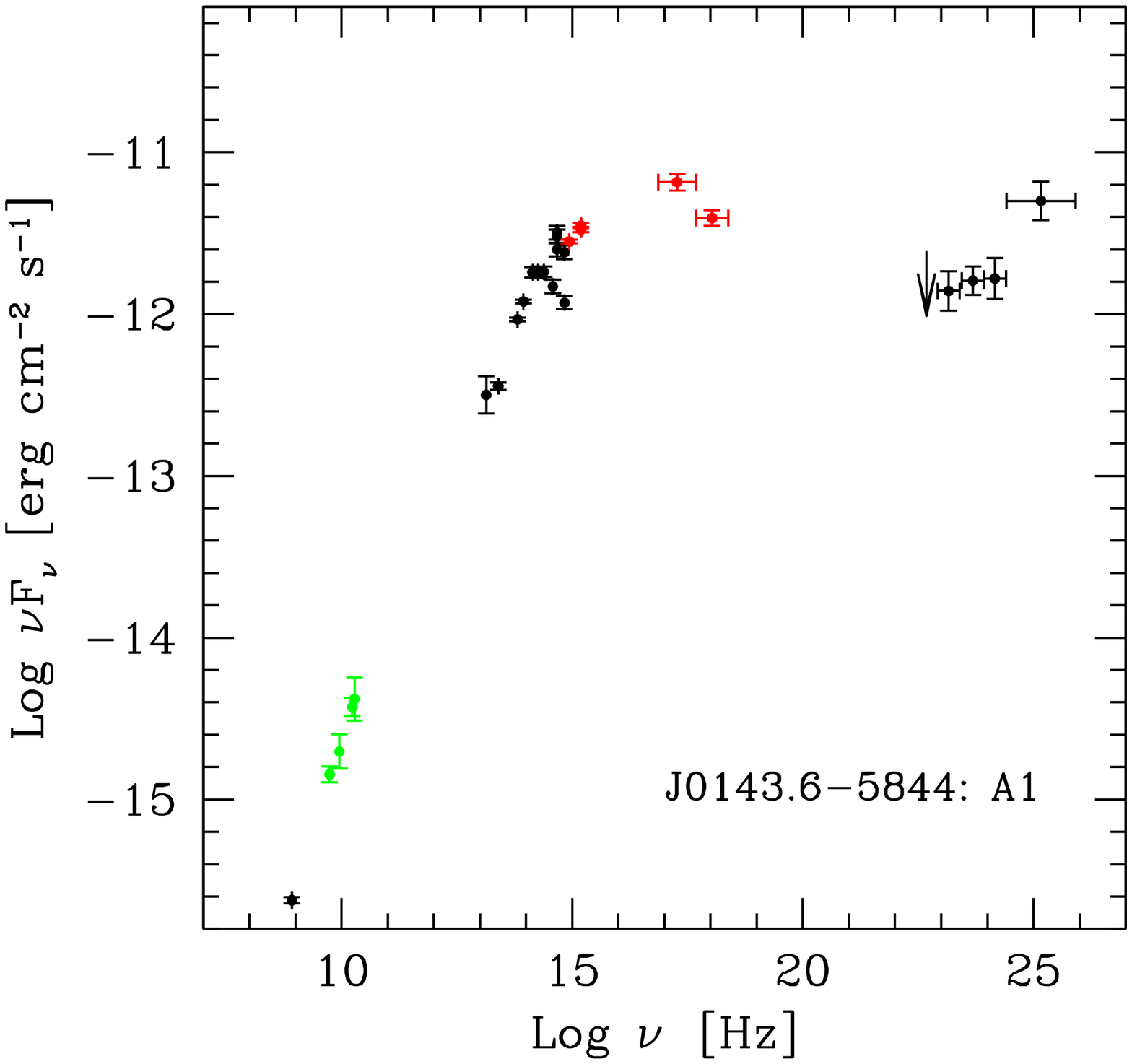} & \hspace{0.4cm}
\includegraphics[width=6cm,bb=23 188 558 680,clip]{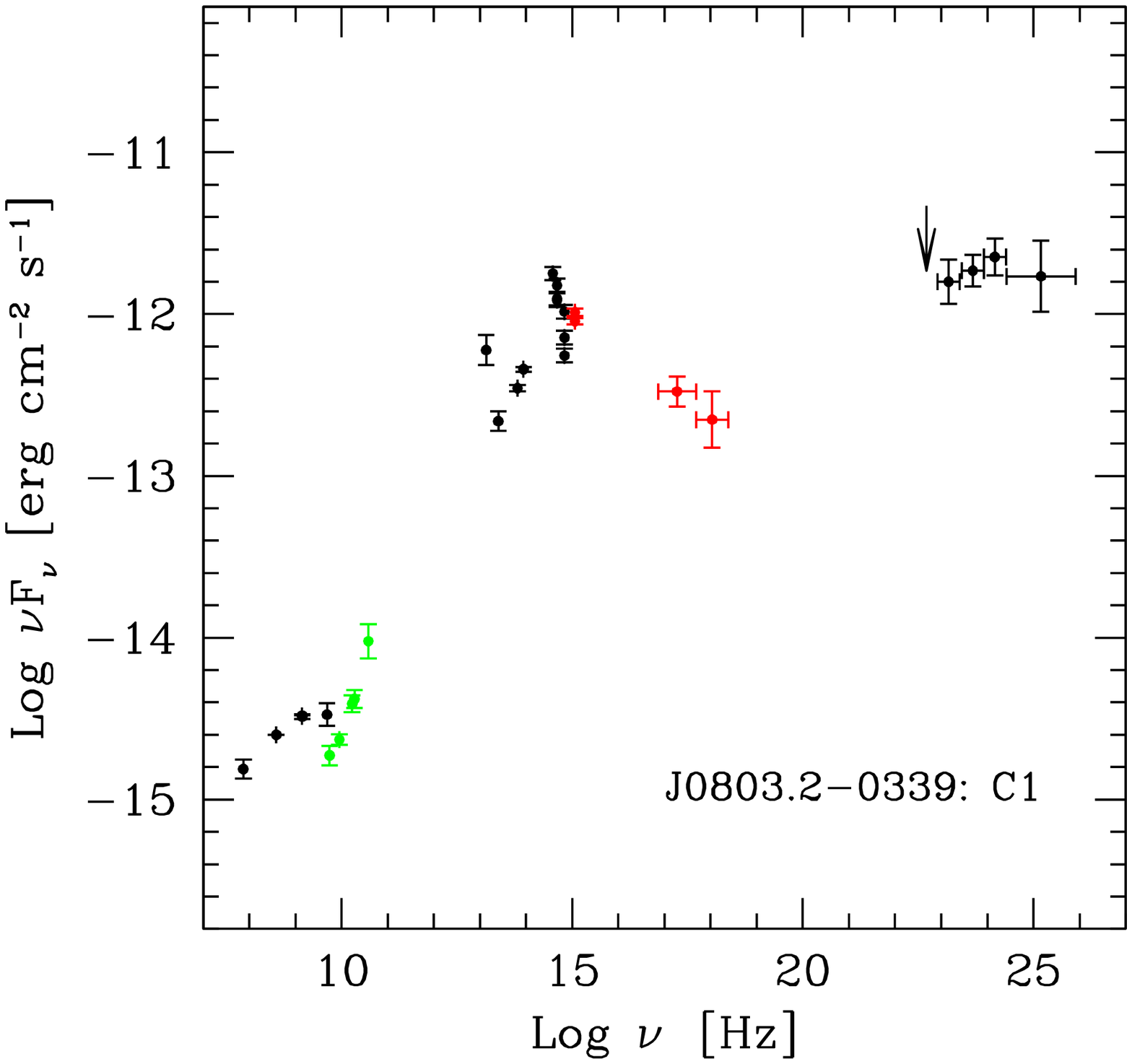} \\

\includegraphics[width=6cm,bb=23 188 558 680,clip]{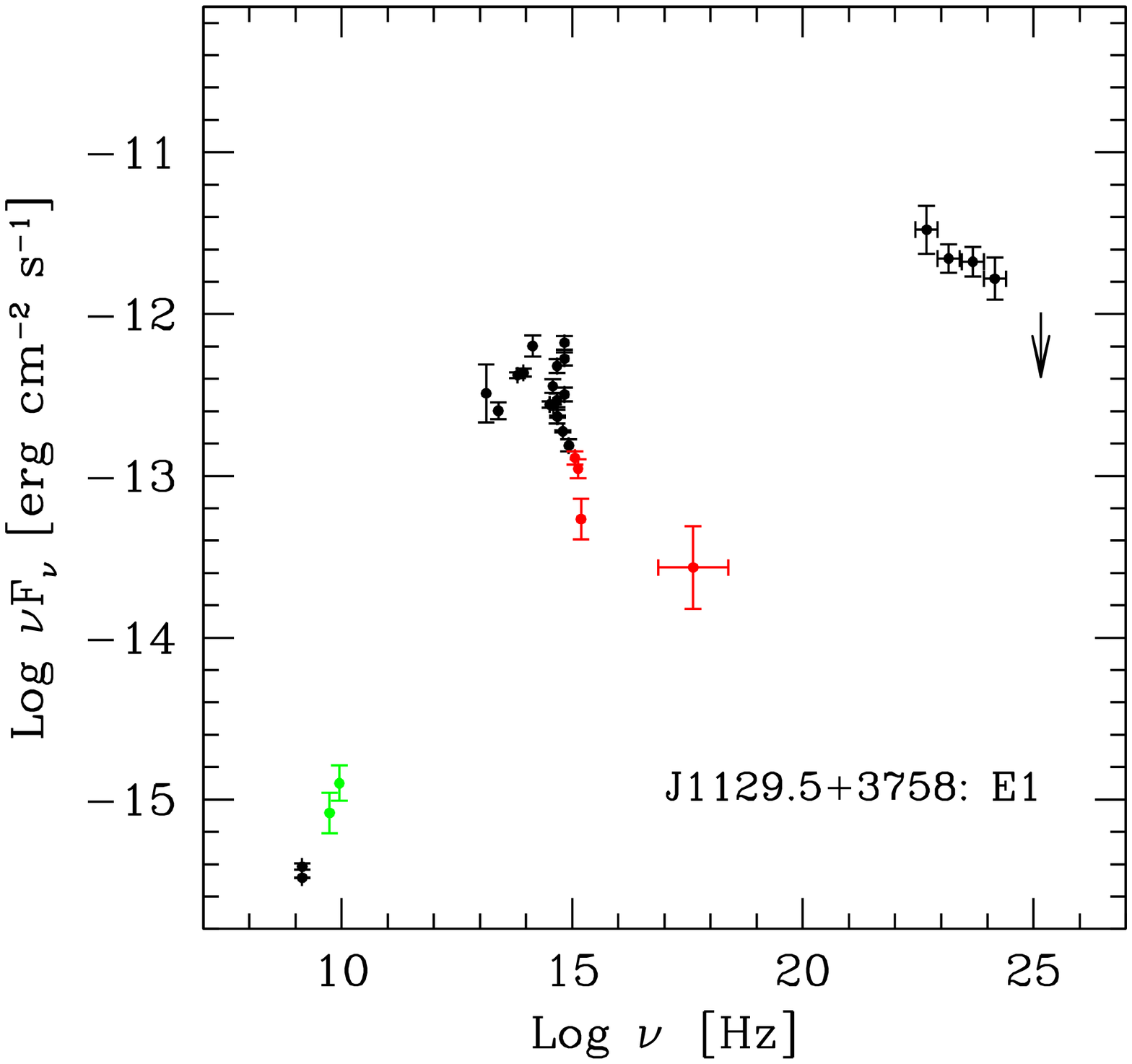} & \hspace{0.4cm}
\includegraphics[width=6cm,bb=23 188 558 680,clip]{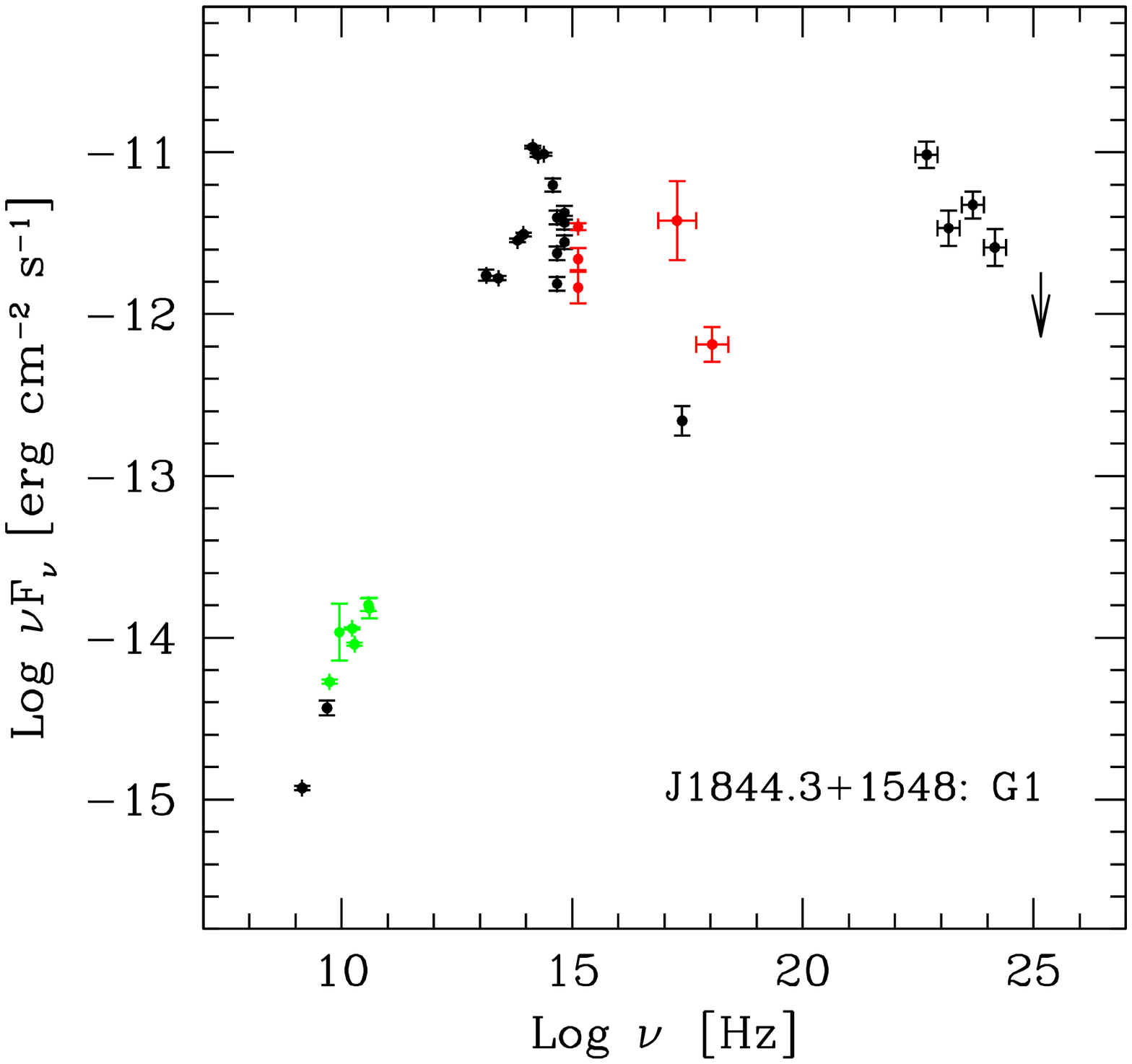}

   \end{tabular}

\caption{Spectral energy distributions of the likely counterpart to the \g-ray sources J\,0143.6$-$5844, J\,0803.2$-$0339, J\,1129.5$+$3758 and J\,1844.3$+$1548 (from top left to bottom right). The green and red 
points corresponds to the ATCA and {\it Swift} data presented in this paper. The black  points represent archival data and are not necessarily simultaneous.}
\label{SED}
\end{figure*}

In order to discuss the multi-wavelength properties of the counterparts presented in the previous section, 
spectral energy distributions (SEDs) were built
for the \g-ray sources where a  radio counterpart was detected in at least two frequencies 
(2FGL J\,0143.6$-$5844, J\,0803.2$-$0339, J\,1129.5$+$3758, and J\,1844.3$+$1548), making these 
sources strong candidates to be AGN.
To generate the SEDs for each candidate we used the ASDC SED Builder \citep{stratta11},
 a web-based program developed at the ASI Science
Data Center that combines data from several missions and experiments, together 
with catalogs, archival and proprietary data. We used fluxes from our radio
campaign, as well as from the UVOT images and the XRT events. The optical/UV 
fluxes were corrected for Galactic extinction, while we used the unabsorbed 
X-ray flux in the 0.3-1.5 and 1.5-10 keV energy ranges (or 0.3-10 keV only for
sources with a limited number of counts).
{In the case of  J\,1129.5$+$3758, where no spectral information could be derived because of the low statistics,
the X-ray flux was derived assuming an absorbed power law with an index of 2.0 (similar to what is observed for the other AGN candidates)
 and the value of the Galactic $n_{\rm H}$ along the line of sight was used.}
 The SEDs are shown in Fig. \ref{SED} and are discussed individually below.

It is interesting to note that \citet{mirabal12} assign 
probabilities of 0.962 and higher for these four objects to be AGN.
{Using an improved version of their infrared colors prediction method, a list of new $\gamma$-ray AGN candidates 
is presented in \citet{massaro13}.  When a prediction is available, a comparison of the results is presented. 
}
%The estimate of the figure of merit (FoM), following a procedure similar to that
%used for associating EGRET and LAT blazars with radio counterparts 
%(Sowards-Emmerd et al. 2003, Abdo et al. 2009), has been obtained a follows:
%.....

%\begin{itemize}

 \medskip

{ \bf  J\,0143.6$-$5844}: Among these 4 sources, the SED of A1 (Fig. \ref{SED}, top-right panel) 
most closely resembles what has been observed in other blazar candidates: there is a
flat radio spectrum that rises to a peak in the UV/soft X--ray range and 
declines in the 
X--rays. This might be interpreted as the synchrotron component of the emission
observed in the high--peaked (BL Lacertae) blazars. The flat spectrum observed
in the {\it Fermi}--LAT data suggests that the high--energy photons belong to a 
different emission process, likely inverse Compton upscatter of photons either
in the relativistic jets (synchrotron self--Compton) or from outside the jets
(external Compton). 
{Other supporting  evidence is found in \citet{massaro13}, where the infrared source WISE J\,014347.39$-$584551.3 (spatially coincident with the X-ray source A1)  is classified as an AGN and more specifically as a BL Lac candidate. }

 \medskip

{ \bf  J\,0803.2$-$0339}: The combination of archival and new radio data might indicate that the 
source C1 has evolved with time, moving from a flat to a steep radio spectrum. However these data 
are not simultaneous and should be interpreted with caution.  This kind of behavior has been seen 
in other blazars \citep[e.g., PKS\,0521$-$36,][]{tornikoski02} when the steep spectrum is seen during 
an active period. Furthermore, the observations of a steep radio spectrum in a quasar by LAT is not 
uncommon \citep{abdo10-agn}. The emission from radio to X--rays is again dominated by synchrotron 
radiation, although in the optical a sign of the contribution from the host galaxy is evident. Also in this case, 
the high-energy photons do not seem to be due to the same radiation mechanism and may be produced
by inverse Compton up-scattering. In addition, we note that the variability index reported in the 2FGL 
catalog for this source \citep[TS$_{\rm var}=48.58$,][]{nolan12}
indicates a \g-ray variability with a $> 99\%$ confidence level\footnote{In the analysis presented in the catalog, a TS$_{\rm var}>41.60$ is used to identify variable sources at a 99\% confidence level.}, supporting an AGN nature for this object. 
{For this source, no $\gamma$-ray AGN candidate was found in \citet{massaro13}.}

\medskip

{\bf J\,1129.5$+$3758}:  The SED of E1 is very puzzling: while the galaxy contribution is an 
evident feature in the near--IR to the UV, the radio, X--ray and \g-ray 
emissions are more difficult to reconcile. The relatively steep spectrum in 
the GeV range and the very low value of the X--ray flux indicate that a peak of
emission must be found between these two ranges. Considering that the hardness 
ratio in the X--rays is $-0.3$, this might indicate a somewhat flat spectrum and, 
consequently, that the X--ray emission might be produced by the same mechanism 
as the \g-rays. Unfortunately the lack of a detailed X--ray spectrum does not 
allow us to draw a firm conclusion. In any case, the parabola that would fit the
inverse Compton emission in the X--rays and the \g-rays would be very narrow. 
The radio emission, meanwhile, cannot be an extrapolation of the higher energy 
emission. If this counterpart is indeed a blazar, then the radio emission might 
be produced by synchrotron with a very low peak emission, in particular when 
compared to the inverse Compton peak. The ratio between the synchrotron and the inverse 
Compton peak might differ by a 
factor of 2-3 orders of magnitude, a relatively uncommon feature among blazars
\citep[see, e.g., ][]{giommi12}.
We also note that there is a hint of $\gamma$-ray variability in the aperture photometry light curve provided by the {\it Fermi} Science
Support Center\footnote{\scriptsize \url{http://fermi.gsfc.nasa.gov/ssc/data/access/lat/2yr\_catalog/ap\_lcs.php}}. 
However this variability is probably contamination
from the nearby (angular separation of 1.7\deg) flaring \g-ray source 2FGL J\,1127.6$+$3622.

\medskip

{\bf J\,1844.3$+$1548}:  The SED of G1 is another intriguing case: the near--IR to the UV is 
thermal radiation produced by the stars in the host galaxy. The very steep X--ray
spectrum is a feature observed in narrow--line Seyfert 1 galaxies \citep[NLSy1, see, 
e.g., ][ and references therein]{grupe04}. In those objects, the steep 
spectrum may be due to intense soft X--ray flux cooling the accretion disk 
corona \citep{maraschi97}. The soft flux is not observed in this source
because of the relatively high intrinsic absorption found in the X--ray 
analysis. If G1 is really a NLSy1, than it belongs to the rare class of 
radio--loud NLSy1 galaxies, a class that have been found to be a \g-ray
emitter \citep{abdo09-agn,foschini11,dammando12}.
In these sources, the emission in the radio and the \g-ray is produced by a blazar-like relativistic jet that is 
dissipating most of its energy beyond the broad-line region \citep{ghisellini08}.
{We note that the infrared source WISE J\,184425.36+154645.9 (spatially coincident with the X-ray source G1)
 is also classified as a BL Lac candidate in \citet{massaro13}.}

\subsection{ Pulsar candidate}

{\bf J\,1036.1$-$6722}: For this object, only two X-ray sources with 
significance $ > 3\sigma $ are detected in a 35-ksec observation, the deepest 
for our sample. Counterparts in the radio were found for only one X-ray source 
(D1) and was only detected at 5.5 GHz, indicating a steep radio spectrum.
In $\gamma$-rays, the best-fit spectral model reported in the 2FGL is a LogParabola.
{In addition, no infrared AGN candidates are reported in \citet{massaro13}  for this $\gamma$-ray source. }
Those characteristics are reminiscent of what is seen for the pulsar/MSP population 
observed with {\it Fermi}-LAT. This \g-ray source is therefore a prime candidate to 
search for \g-ray pulsation at the position of the newly detected counterpart.
The \textit{much} reduced position error ellipse (from 0.1\deg to $\sim$4'') 
will allow exploration of the parameter space (e.g., pulsar period) 
in considerably greater detail. We note that in the case of a binary system, the measurement of
the orbital period (through e.g., optical observations) will greatly enhance the probability of detecting the pulsation.

\subsection{ Intriguing objects }

{\bf J\,0523.3$-$2530}: Although J\,0523.3$-$2530 is the brightest \g-ray source in 
our sample, only one faint X-ray source was detected, with a radio counterpart 
only at the lowest radio frequency. While the radio, UV, and X-ray observations might 
suggest a pulsar origin, the \g-ray spectrum is best represented by a power-law
as seen for other AGN candidates. {However, \citet{massaro13} report no AGN candidate for this source and interestingly, 
the probability to be an AGN is P(AGN)=0.738 \citep{mirabal12}, a value that places this source neither in 
the pulsar nor in the AGN class.}

\medskip

{\bf J\,1231.3$-$5112}: The two interesting X-ray counterparts (F1 and F2) have dim UV
counterparts and no detection in the radio. This unidentified source has 
intriguing \g-ray properties, as the best-fit spectral model, a LogParabola, may
suggest a pulsar origin while the light curve is not constant, with a \g-ray 
variability index of \citep[TS$_{\rm var}=39.04$,][]{nolan12}. As for the
previous source, the probability to be an AGN is at an intermediate value, 
P(AGN)=0.554.

\medskip

Further radio and X-ray follow-up observations of these sources are therefore 
required to understand their nature.

\subsection{X-ray chance coincidence for non-AGN candidates}

%{For the AGN candidates, the detection of a flat spectrum in radio and X-ray variability provide strong evidence that the considered X-ray source 
%is the likely counterpart to the $\gamma$-ray source. However, for non-AGN candidates (pulsar and intriguing sources) this is different as few 
%or no radio information was gathered. For this category of sources, the probability of chance coincidence was investigated.
%  }

Although the X-ray to radio connection, combined with the detection of X-ray variability and the presence of a flat radio spectrum, is a step forward in pinpointing AGNs, a simple detection in the optical/UV/X-ray regimes is  insufficient in determining the correct counterpart for pulsar-like $\gamma$-ray sources. 
For this category of sources, the probability of chance coincidence was investigated.
Among the 3 non-AGN candidates, two of them (J\,0523.3$-$2530 and J\,1231.3$-$5112) had no \textit{Swift}  XRT detection outside the R$_{95}$ error ellipse.
The X-ray sources presented in Table \ref{table} are therefore promising counterpart candidates.
For J\,1036.1$-$6722, whose \textit{Swift} exposure time was longer (35.5 ksec, compared to a standard $\sim$4 ksec for the other sources), 
13 other sources were observed outside the R$_{95}$ region.
The majority (8) of these sources have a detection significance just above 3$\sigma$ and no additional information can be obtained.
We tentatively tried to look for peculiar behaviors for the remaining 5 sources, whose detection significance is in the range 4.5-6.3$\sigma$, using the XRT and the UVOT data. The analysis of the 26 XRT observations does not show any sign of significant variability (above 1$\sigma$).  Inspection of the 4 UVOT filters shows that of the 5 X-ray sources, one does not have any optical/UV counterpart, 2 have very weak counterparts, while the other 2 have strong counterparts. Unfortunately, we are not able to estimate the variability of the latter 2 sources, since they are located outside
 the UVOT FoV\footnote{The UVOT FoV is 17'x17', while the XRT camera has a circular shape of 12' in radius.} most of the time. 
We estimated the chance coincidence of finding X-ray sources within the R$_{95}$ region following the method explained in \citet{bloom02}.
 The probability of chance association P can be expressed as $P = 1 - e^{- A \times \rho}$
where A is the area of the R$_{95}$ region (a $0.062\deg \times 0.058\deg$ ellipse), while $\rho$ is the sky density of objects with equal or greater X-ray brightness. Since there are 15 sources in the XRT FoV (a circular shape 12' in radius) detected above the 3$\sigma$ level (D2 in Table \ref{table}), the chance probability is 0.74.

\textbf{While this value is high, D1 has a slightly higher chance to be the \g-ray counterpart than other X-ray detected sources because it is located within the R$_{95}$ region, is the second brightest object in the XRT field, and is detected at 5.5 GHz. However,  other sources detected in the XRT field can still be considered as candidates for the association. }

%%%%%%%%%%%%%%%%%%%%%%%%%%%%%%%%%%%%%%%%%%%%%%%%%%%%%%%%%%%%%%%%%%%%%%%%%%%%%%%{\it Fermi}
\section{Summary and conclusion}

%%%%%%%%%%%%%%%%%%%%%%%%%%%%%%%%%%%%%%%%%%%%%%%%%%

We have used the combination of X-ray and radio follow-up observations to 
investigate the nature of a sample of bright, high latitude 2FGL unassociated 
sources. X-ray observations from the \textit{Swift} satellite were used to 
identify potential counterparts within the {\it Fermi} position error ellipse, which were 
then targeted with radio follow-up observations. The nature of the counterparts and their
possible association with the \g-ray source was then discussed in a multi-wavelength context.
Out of the 7 objects presented in this study, 4 \g-ray sources show a converging
trend of evidence that suggests an AGN nature. Among those sources, J\,1844.3$+$1548
could be associated with a  narrow--line Seyfert 1 galaxy, a rare class of 
\g-ray emitter. Although the limited sample considered here, we note that X-rays are good tracers of potential AGN 
counterparts as, when a bright X-ray source was found in our sample, a radio 
flat spectrum counterpart was detected. Follow-up optical observations are 
planned to determine the redshift of those AGN candidates as well as to confirm 
the narrow--line Seyfert 1 nature of J\,1844.3$+$1548. Our study confirms that the 
combination of X-rays and radio follow-up observations provides an efficient 
method to identify AGN candidates among the {\it Fermi} unassociated sources.
It is also interesting to note that our list of AGN candidates constructed using
multi-wavelength properties is in agreement with the prediction, based on the GeV properties
only, made by \citet{mirabal12}. 

The identification of candidates belonging to known classes of astrophysical objects 
narrows down the list of associated sources to the truly exotic objects. Additionally, the 
newly discovered members of known classes are also of great interest as they represent 
outliers in their respective classes. For example, our new candidate AGN (J\,1129.5$+$3758) may belong to a
potentially important subclass: high Compton dominance AGN. Compton dominance is
the ratio of the peak Compton to the peak synchrotron luminosity. The 
radio-weak, \g-ray loud AGN we detect are increasing the size of this key sample,
which can address the many questions that the relation of the Compton dominance
to AGN properties like peak synchrotron frequency \citep{finke13} 
raise, e.g., the existence of the ``blazar sequence" \citep{fossati98}.

%The brightest sources in our sample turned out to be interesting pulsar candidate and the precise location of counterparts will allow deeper pulsation searches in the \g-rays.
The precise location of the putative counterpart of the pulsar candidate we have identified 
in our sample (J\,1036.1-6722), will be used to perform deep \g-ray pulsation searches. The 
improved location translates to a dramatic reduction in the volume of phase space that will 
need to be searched, making this task computationally more feasible. 

In \citet{su12}, the line at 130 GeV (potentially resulting from dark matter 
annihilation) is investigated in a sample of  {\it Fermi} unassociated sources, of 
which J\,1844.3$+$1548 is the brightest source. In this work we have demonstrated 
that this object is probably associated with an AGN and unlikely to be a 
signature of dark matter annihilation.  

In our hunt to identify the truly ``exotic" objects, we found an intriguing 
pair (J\,0523.3-2530 and J\,1231.3-5112) of \g-ray sources whose multi-wavelength properties do 
not seem to fit  in the pulsar or in the AGN category. Further investigation of
those extra-ordinary sources could provide a pathway to the discovery of new 
types of \g-ray emitters.

%%%%%%%%%%%%%%%%%%%%%%%%%%%%%%%%%%%%%%%%%%%%%%%%%%%%%%%%%%%%%%%%%%%%%%%%%%%%%%%

%%%%%%%%%%%
\begin{figure*}
 \centering

   \begin{tabular}{ccc}

%\hspace{-0.5cm}
\includegraphics[width=6cm,bb=40 170 566 629,clip]{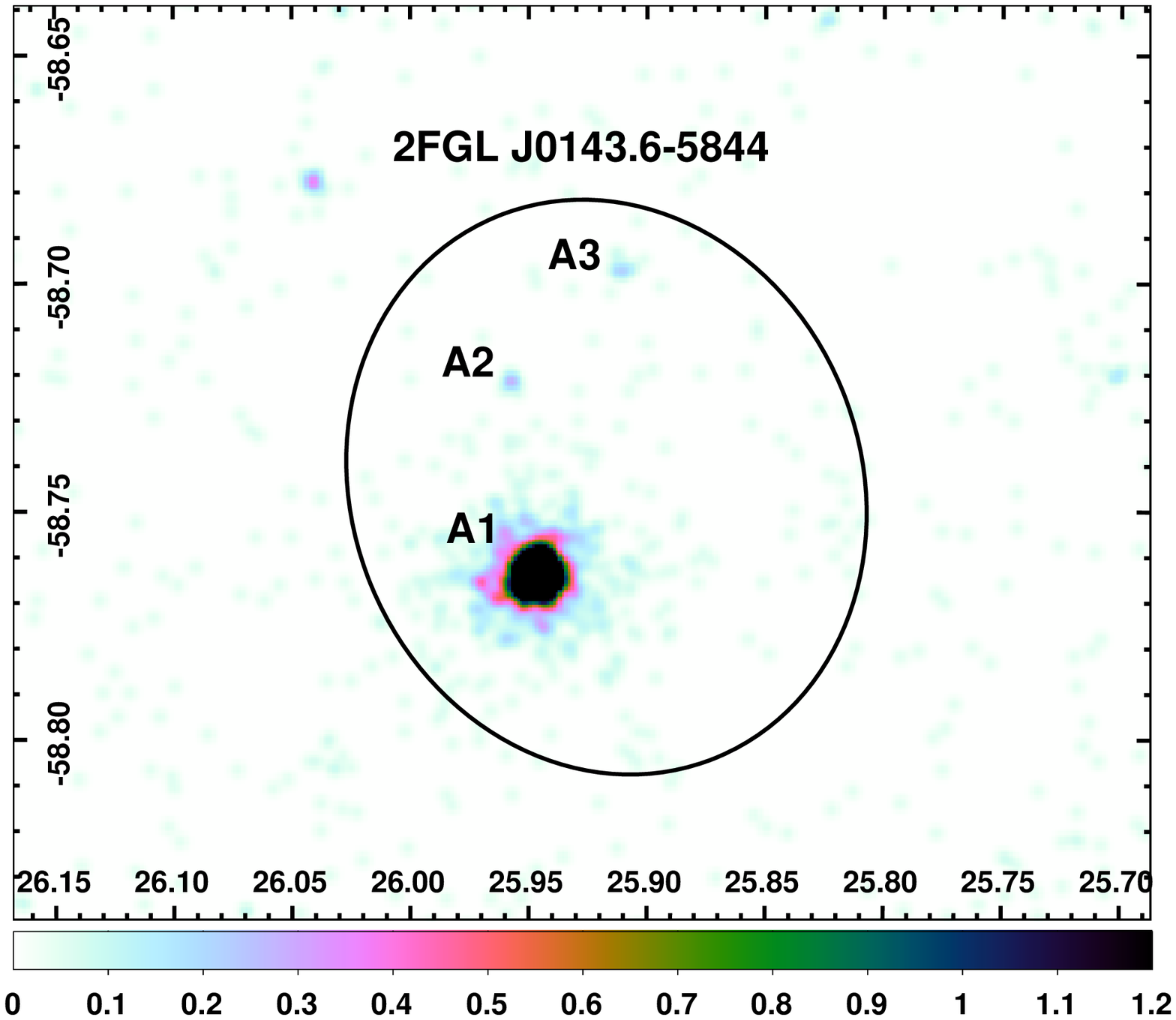} & \hspace{-0.4cm}
\includegraphics[width=6cm,bb=40 170 566 629,clip]{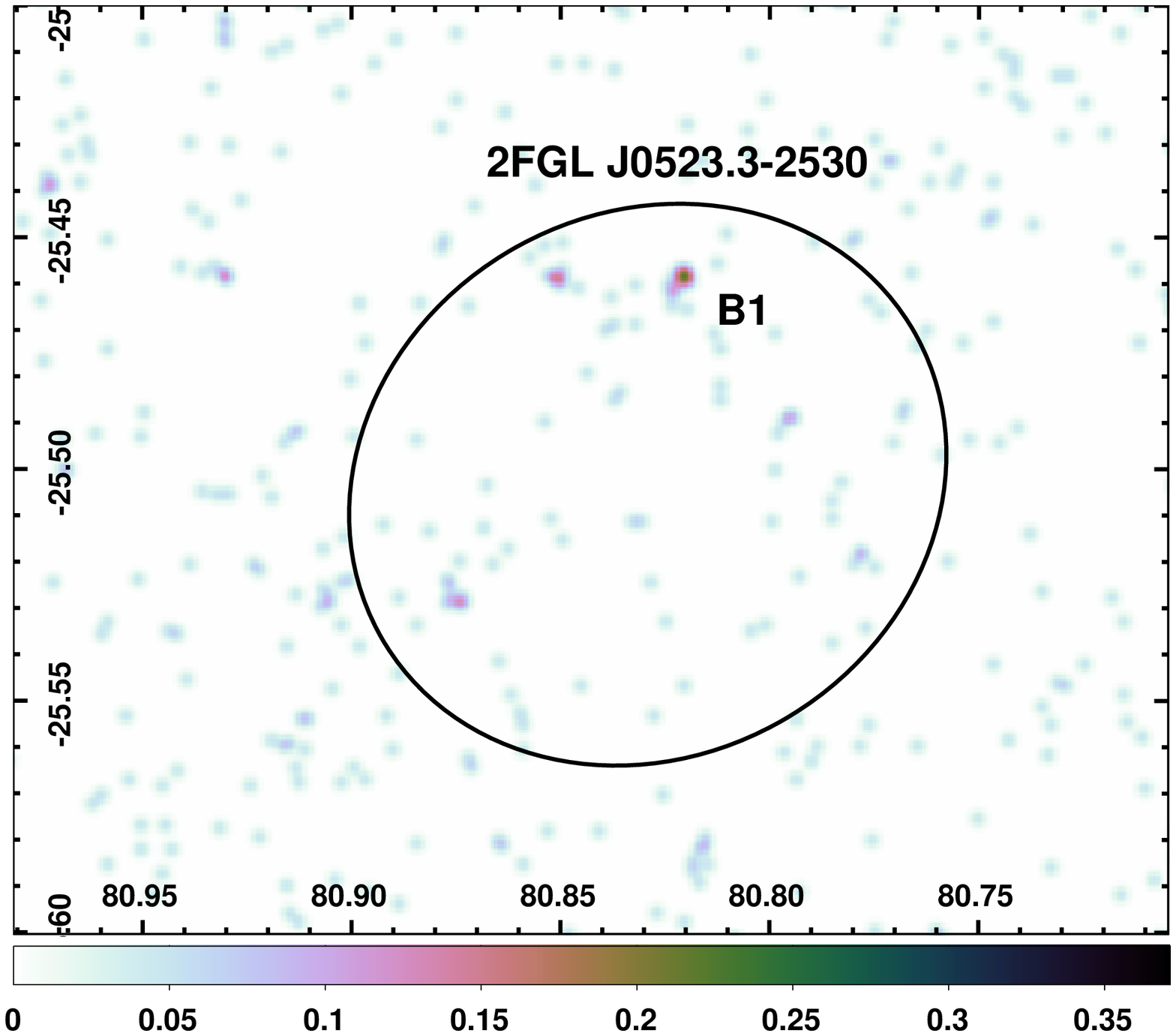} & \hspace{-0.4cm}
\includegraphics[width=6cm,bb=40 170 566 629,clip]{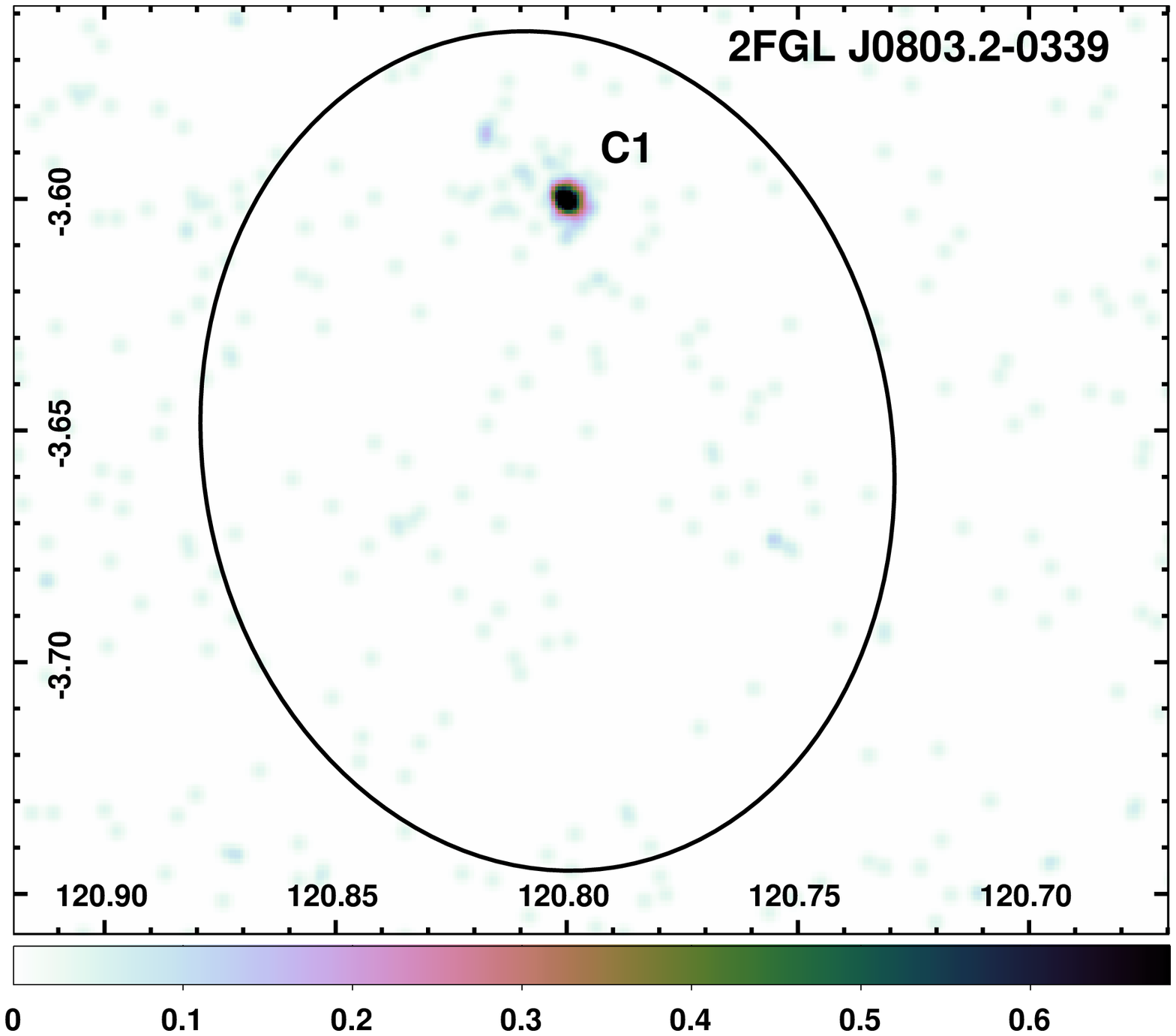} \\

%\vspace{-0.9cm}

\includegraphics[width=6cm,bb=40 170 566 629,clip]{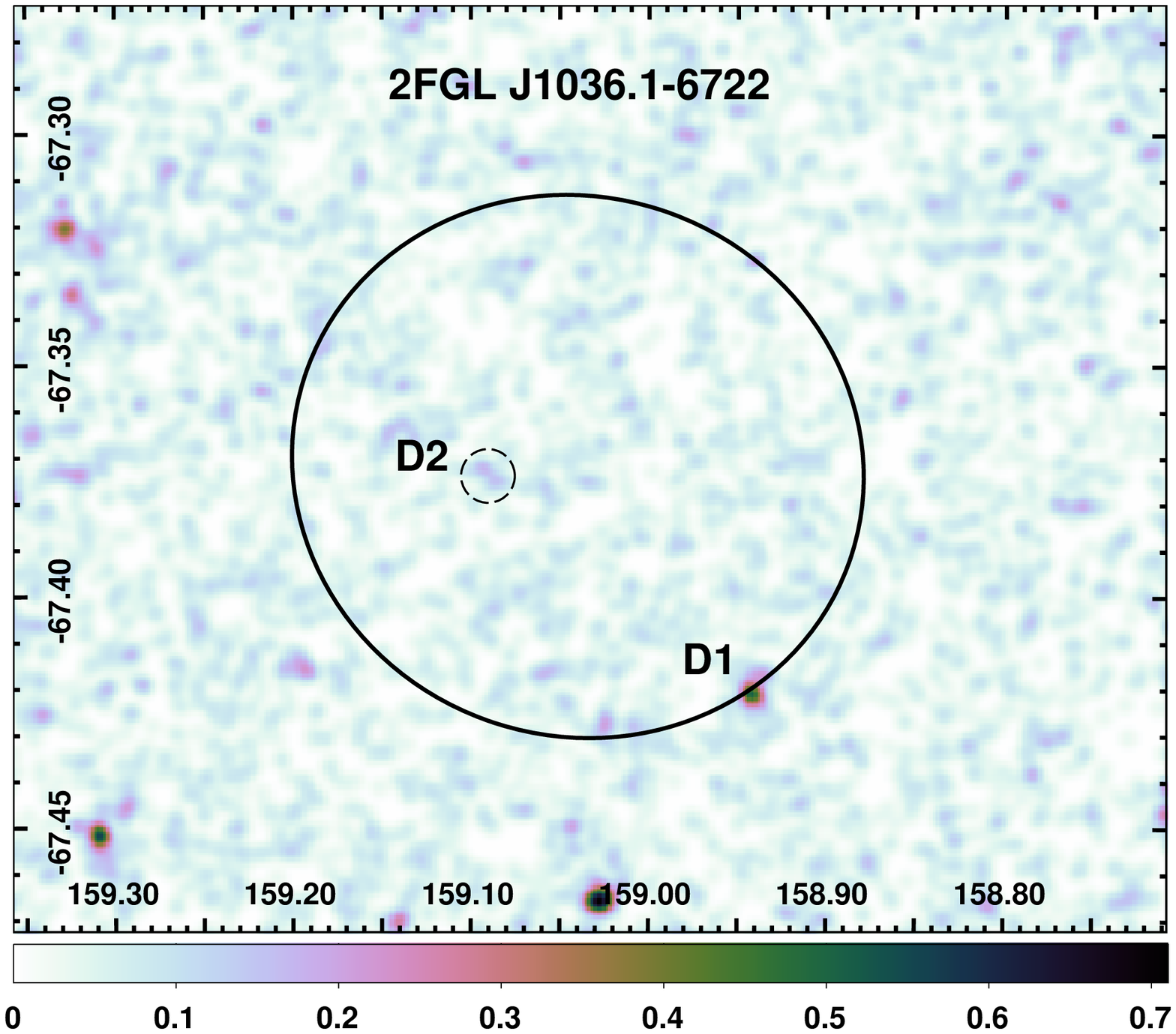} & \hspace{-0.4cm}
\includegraphics[width=6cm,bb=40 170 566 629,clip]{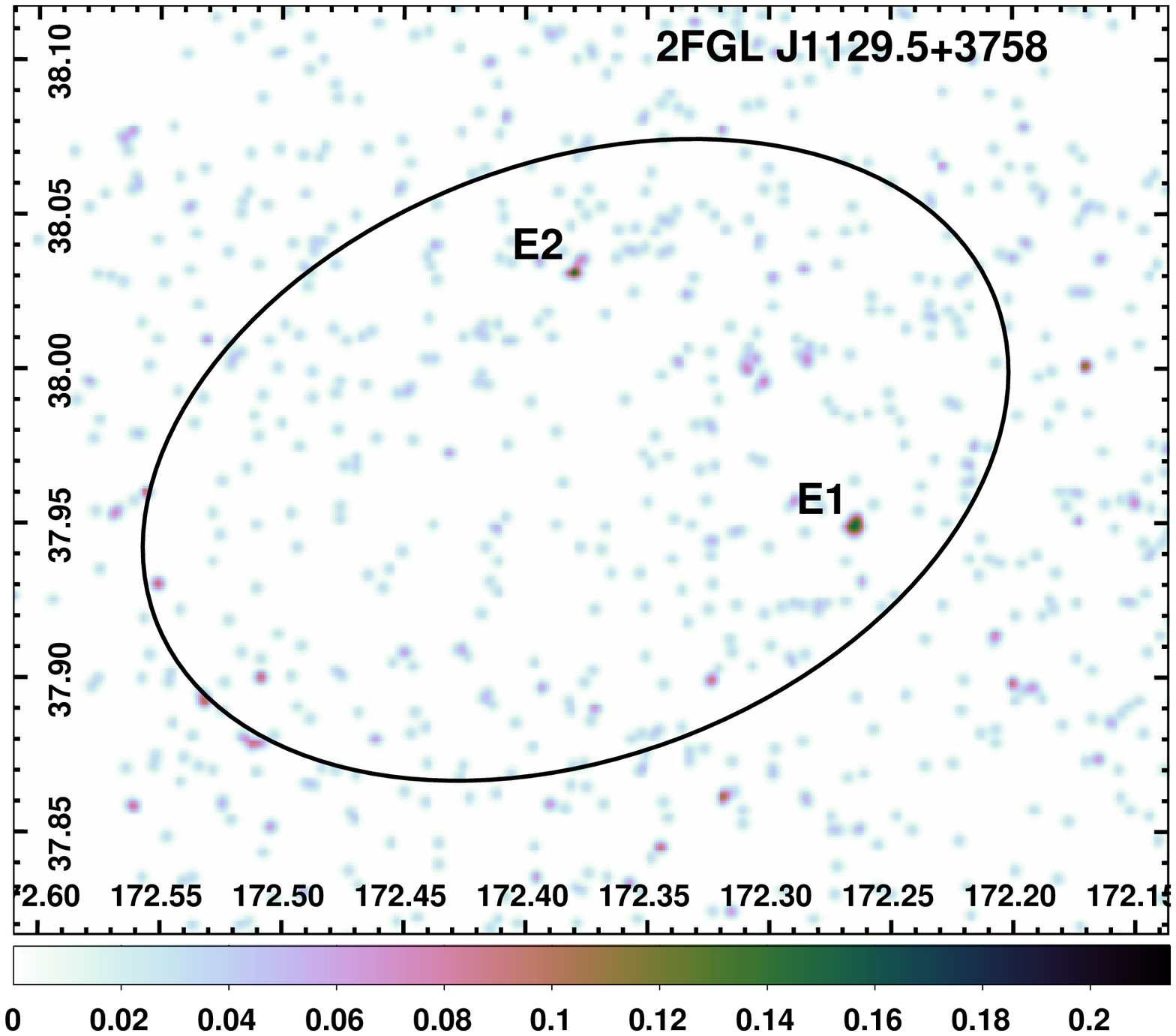} & \hspace{-0.4cm}

\includegraphics[width=6cm,bb=40 170 566 629,clip]{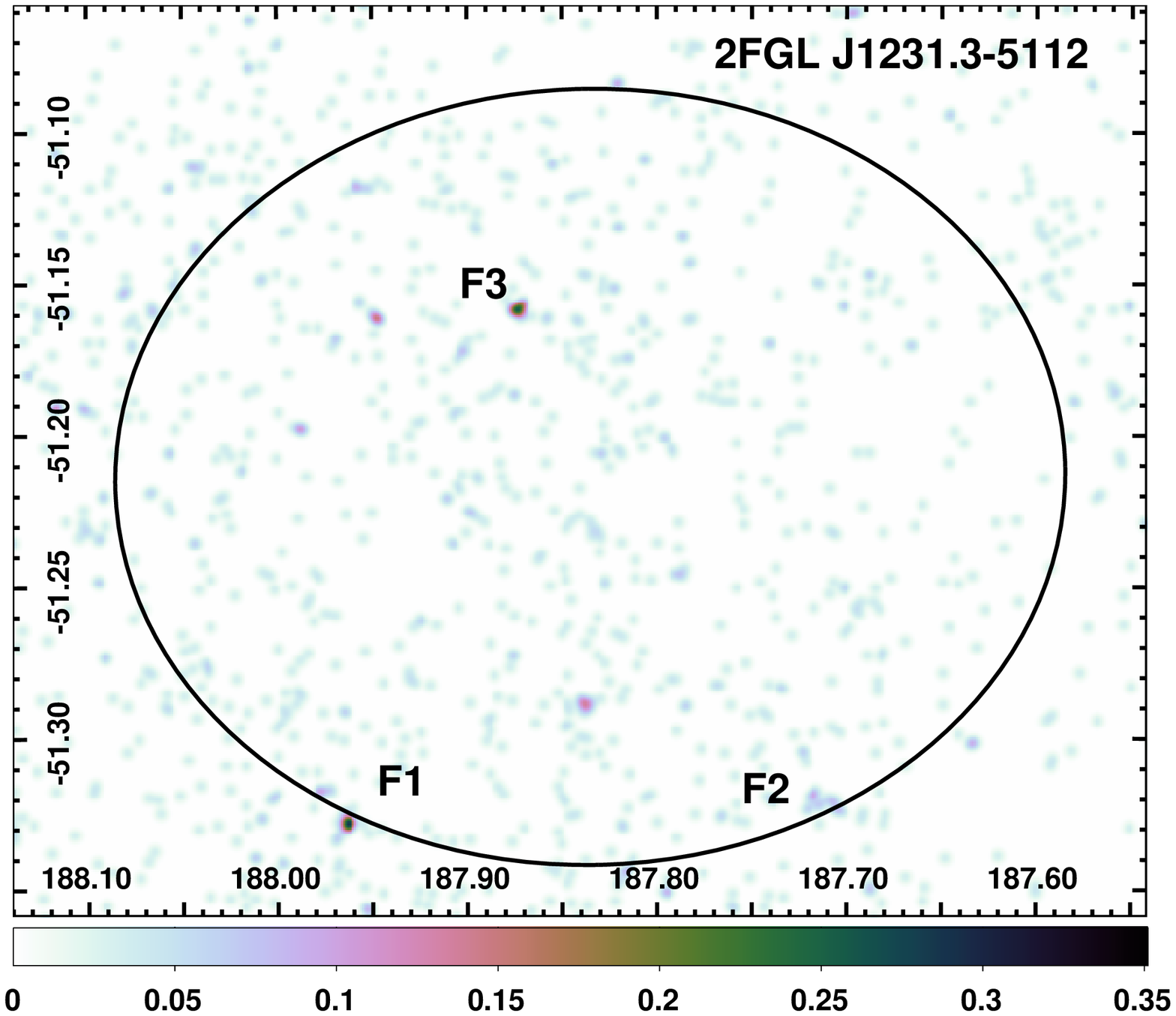}  \\

\includegraphics[width=6cm,bb=40 170 566 629,clip]{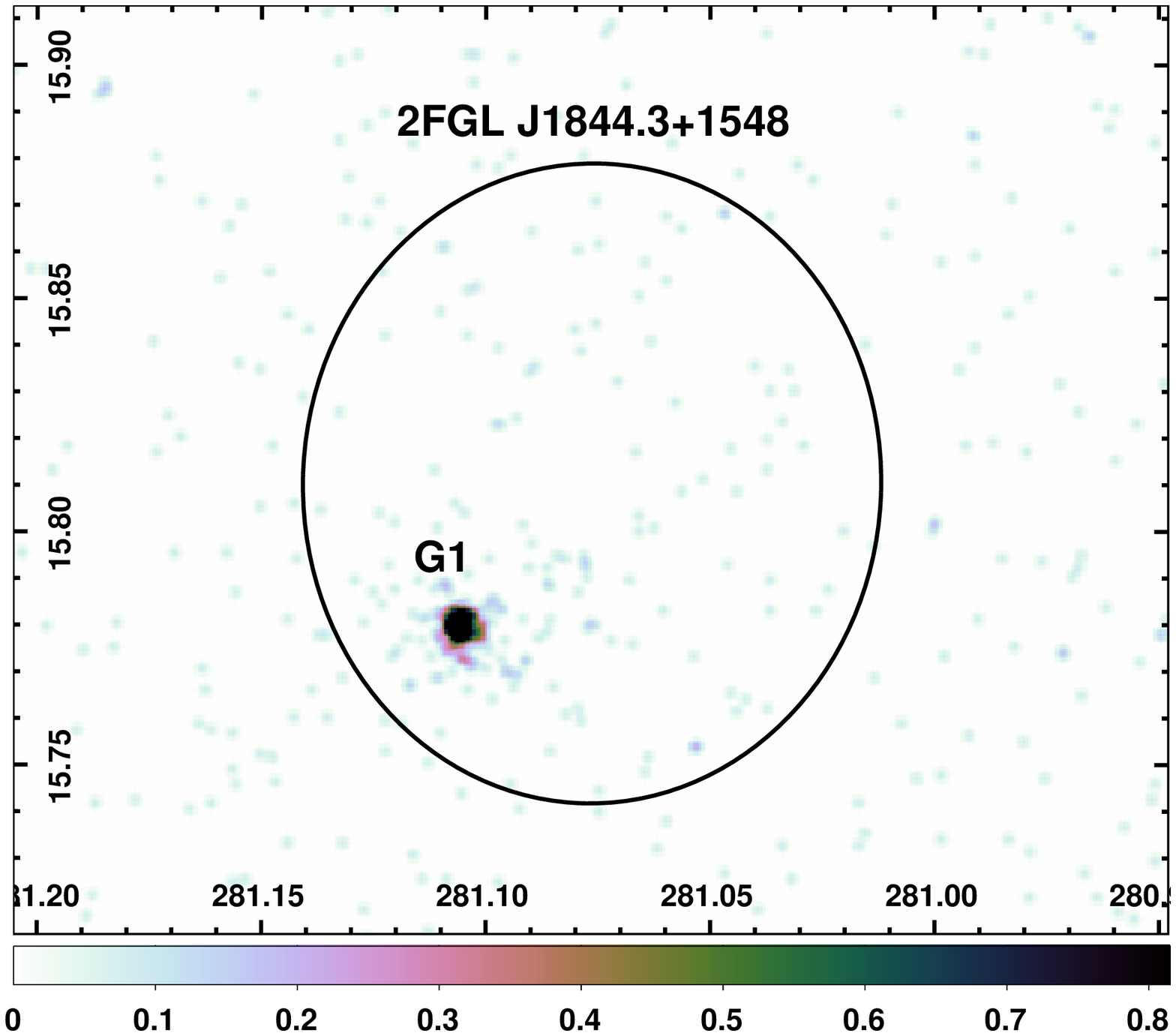} \\

   \end{tabular}

\caption{{\it Swift} X-ray count maps (0.3-10 keV energy band) of the selected 
sample of 2FGL unassociated sources with X-ray coverage and dedicated follow up 
radio observations. The black ellipse represent the 95\% confidence level 
position of the {\it Fermi} source. The X-ray sources presented in Table 
\ref{table} are labelled on the images. The images are in RA/DEC coordinate and the color scale is linear. The 
images have been smoothed with a Gaussian of 7\arcsec   kernel. }
\label{image}
\end{figure*}

%%%%%%%%%%%%%%%%%%%%%%%%%%%%%%%%%%%%%%%%%%%%%%%%%%%%%%%%%%%%%%%%%%%%%%%%%%%%%%%

%%%%%%%%%%%

\acknowledgments
This work made use of data supplied by the UK \textit{Swift }Science Data Centre at the University of Leicester \citep{evans09}. 
The Australia Telescope Compact Array is part of the Australia Telescope National Facility which is funded by the Commonwealth of Australia for operation as a National Facility managed by CSIRO.
This research was funded in part by NASA through {\it Fermi} Guest Investigator grants NNH09ZDA001N and NNH10ZDA001N. This research was supported by an appointment to the NASA Postdoctoral Program at the Goddard Space Flight Center, administered by Oak Ridge Associated Universities through a contract with NASA. 
This research has made use of data from the NASA/IPAC
Extragalactic Database (NED, operated by the Jet Propulsion Laboratory,
California Institute of Technology, under contract with the National
Aeronautics and Space Administration); and the SIMBAD database (operated
at CDS, Strasbourg, France). This research has made use of NASA's
Astrophysics Data System.

{\it Facilities:} \facility{ATCA}, \facility{Fermi},  \facility{Swift}.

\bibliographystyle{mn2e}

\begin{thebibliography}{27}
\expandafter\ifx\csname natexlab\endcsname\relax\def\natexlab#1{#1}\fi

\bibitem[{{Abdo} {et~al}\mbox{.}(2010{\natexlab{a}}){Abdo}, {Ackermann},
  {Ajello}, {Atwood}, {Axelsson}, {Baldini}, {Ballet}, {Barbiellini}, {Baring},
  {Bastieri}, \& et~al.}]{abdo10-1pc}
{Abdo} A.~A. {et~al.}, 2010{\natexlab{a}}, \apjs, 187, 460

\bibitem[{{Abdo} {et~al}\mbox{.}(2010{\natexlab{b}}){Abdo}, {Ackermann},
  {Ajello}, {Baldini}, {Ballet}, {Barbiellini}, {Bastieri}, {Bechtol},
  {Bellazzini}, {Berenji}, {Blandford}, {Bloom}, {Bonamente}, {Borgland},
  {Bouvier}, {Brandt}, {Bregeon}, {Brez}, {Brigida}, {Bruel}, {Buehler},
  {Burnett}, {Buson}, {Caliandro}, {Cameron}, {Cannon}, {Caraveo}, {Carrigan},
  {Casandjian}, {Cavazzuti}, {Cecchi}, {{\c C}elik}, {Celotti}, {Charles},
  {Chekhtman}, {Chen}, {Cheung}, {Chiang}, {Ciprini}, {Claus}, {Cohen-Tanugi},
  {Colafrancesco}, {Conrad}, {Davis}, {Dermer}, {de Angelis}, {de Palma},
  {Silva}, {Drell}, {Dubois}, {Favuzzi}, {Fegan}, {Ferrara}, {Fortin},
  {Frailis}, {Fukazawa}, {Fusco}, {Gargano}, {Gasparrini}, {Gehrels},
  {Germani}, {Giglietto}, {Giommi}, {Giordano}, {Giroletti}, {Glanzman},
  {Godfrey}, {Grandi}, {Grenier}, {Grove}, {Guillemot}, {Guiriec}, {Hadasch},
  {Hayashida}, {Hays}, {Horan}, {Hughes}, {Jackson}, {J{\'o}hannesson},
  {Johnson}, {Johnson}, {Kamae}, {Katagiri}, {Kataoka}, {Kn{\"o}dlseder},
  {Kuss}, {Lande}, {Latronico}, {Lee}, {Lemoine-Goumard}, {Llena Garde},
  {Longo}, {Loparco}, {Lott}, {Lovellette}, {Lubrano}, {Madejski}, {Makeev},
  {Malaguti}, {Mazziotta}, {McConville}, {McEnery}, {Michelson}, {Migliori},
  {Mitthumsiri}, {Mizuno}, {Monte}, {Monzani}, {Morselli}, {Moskalenko},
  {Murgia}, {Naumann-Godo}, {Nestoras}, {Nolan}, {Norris}, {Nuss}, {Ohsugi},
  {Okumura}, {Omodei}, {Orlando}, {Ormes}, {Paneque}, {Panetta}, {Parent},
  {Pelassa}, {Pepe}, {Persic}, {Pesce-Rollins}, {Piron}, {Porter}, {Rain{\`o}},
  {Rando}, {Razzano}, {Razzaque}, {Reimer}, {Reimer}, {Reyes}, {Roth},
  {Sadrozinski}, {Sanchez}, {Sander}, {Scargle}, {Sgr{\`o}}, {Siskind},
  {Smith}, {Spandre}, {Spinelli}, {Stawarz}, {Stecker}, {Strickman}, {Suson},
  {Takahashi}, {Tanaka}, {Thayer}, {Thayer}, {Thompson}, {Tibaldo}, {Torres},
  {Torresi}, {Tosti}, {Tramacere}, {Uchiyama}, {Usher}, {Vandenbroucke},
  {Vasileiou}, {Vilchez}, {Villata}, {Vitale}, {Waite}, {Wang}, {Winer},
  {Wood}, {Yang}, {Ylinen}, \& {Ziegler}}]{abdo10-agn}
{Abdo} A.~A. {et~al.}, 2010{\natexlab{b}}, \apj, 720, 912

\bibitem[{{Abdo} {et~al}\mbox{.}(2009){Abdo}, {Ackermann}, {Ajello}, {Baldini},
  {Ballet}, {Barbiellini}, {Bastieri}, {Bechtol}, {Bellazzini}, {Berenji},
  {Bloom}, {Bonamente}, {Borgland}, {Bregeon}, {Brez}, {Brigida}, {Bruel},
  {Burnett}, {Caliandro}, {Cameron}, {Caraveo}, {Casandjian}, {Cecchi}, {{\c
  C}elik}, {Chekhtman}, {Cheung}, {Chiang}, {Ciprini}, {Claus}, {Cohen-Tanugi},
  {Conrad}, {Cutini}, {Dermer}, {de Palma}, {Silva}, {Drell}, {Dubois},
  {Dumora}, {Farnier}, {Favuzzi}, {Fegan}, {Focke}, {Foschini}, {Frailis},
  {Fukazawa}, {Fusco}, {Gargano}, {Gehrels}, {Germani}, {Giebels}, {Giglietto},
  {Giordano}, {Giroletti}, {Glanzman}, {Godfrey}, {Grenier}, {Grove},
  {Guillemot}, {Guiriec}, {Hayashida}, {Hays}, {Horan}, {Hughes},
  {J{\'o}hannesson}, {Johnson}, {Johnson}, {Kadler}, {Kamae}, {Katagiri},
  {Kataoka}, {Kerr}, {Kn{\"o}dlseder}, {Kuss}, {Lande}, {Latronico}, {Longo},
  {Loparco}, {Lott}, {Lovellette}, {Lubrano}, {Makeev}, {Mazziotta},
  {McConville}, {McEnery}, {Meurer}, {Michelson}, {Mitthumsiri}, {Mizuno},
  {Monte}, {Monzani}, {Morselli}, {Moskalenko}, {Murgia}, {Nolan}, {Norris},
  {Nuss}, {Ohsugi}, {Omodei}, {Orlando}, {Ormes}, {Pelassa}, {Pepe}, {Persic},
  {Pesce-Rollins}, {Piron}, {Porter}, {Rain{\`o}}, {Rando}, {Razzano},
  {Rochester}, {Rodriguez}, {Ryde}, {Sadrozinski}, {Sambruna}, {Sander}, {Saz
  Parkinson}, {Scargle}, {Sgr{\`o}}, {Smith}, {Spandre}, {Spinelli},
  {Strickman}, {Suson}, {Tagliaferri}, {Takahashi}, {Takahashi}, {Tanaka},
  {Thayer}, {Thayer}, {Thompson}, {Tibaldo}, {Tibolla}, {Torres}, {Tosti},
  {Tramacere}, {Uchiyama}, {Usher}, {Vasileiou}, {Vilchez}, {Vitale}, {Waite},
  {Wang}, {Winer}, {Wood}, {Ylinen}, {Ziegler}, {Fermi/LAT Collaboration},
  {Ghisellini}, {Maraschi}, \& {Tavecchio}}]{abdo09-agn}
{Abdo} A.~A. {et~al.}, 2009, \apjl, 707, L142

\bibitem[{{Ackermann} {et~al}\mbox{.}(2011){Ackermann}, {Ajello}, {Allafort},
  {Antolini}, {Atwood}, {Axelsson}, {Baldini}, {Ballet}, {Barbiellini},
  {Bastieri}, {Bechtol}, {Bellazzini}, {Berenji}, {Blandford}, {Bloom},
  {Bonamente}, {Borgland}, {Bottacini}, {Bouvier}, {Bregeon}, {Brigida},
  {Bruel}, {Buehler}, {Burnett}, {Buson}, {Caliandro}, {Cameron}, {Caraveo},
  {Casandjian}, {Cavazzuti}, {Cecchi}, {Charles}, {Cheung}, {Chiang},
  {Ciprini}, {Claus}, {Cohen-Tanugi}, {Conrad}, {Costamante}, {Cutini}, {de
  Angelis}, {de Palma}, {Dermer}, {Digel}, {Silva}, {Drell}, {Dubois},
  {Escande}, {Favuzzi}, {Fegan}, {Ferrara}, {Finke}, {Focke}, {Fortin},
  {Frailis}, {Fukazawa}, {Funk}, {Fusco}, {Gargano}, {Gasparrini}, {Gehrels},
  {Germani}, {Giebels}, {Giglietto}, {Giommi}, {Giordano}, {Giroletti},
  {Glanzman}, {Godfrey}, {Grenier}, {Grove}, {Guiriec}, {Gustafsson},
  {Hadasch}, {Hayashida}, {Hays}, {Healey}, {Horan}, {Hou}, {Hughes},
  {Iafrate}, {J{\'o}hannesson}, {Johnson}, {Johnson}, {Kamae}, {Katagiri},
  {Kataoka}, {Kn{\"o}dlseder}, {Kuss}, {Lande}, {Larsson}, {Latronico},
  {Longo}, {Loparco}, {Lott}, {Lovellette}, {Lubrano}, {Madejski}, {Mazziotta},
  {McConville}, {McEnery}, {Michelson}, {Mitthumsiri}, {Mizuno}, {Moiseev},
  {Monte}, {Monzani}, {Moretti}, {Morselli}, {Moskalenko}, {Murgia},
  {Nakamori}, {Naumann-Godo}, {Nolan}, {Norris}, {Nuss}, {Ohno}, {Ohsugi},
  {Okumura}, {Omodei}, {Orienti}, {Orlando}, {Ormes}, {Ozaki}, {Paneque},
  {Parent}, {Pesce-Rollins}, {Pierbattista}, {Piranomonte}, {Piron}, {Pivato},
  {Porter}, {Rain{\`o}}, {Rando}, {Razzano}, {Razzaque}, {Reimer}, {Reimer},
  {Ritz}, {Rochester}, {Romani}, {Roth}, {Sanchez}, {Sbarra}, {Scargle},
  {Schalk}, {Sgr{\`o}}, {Shaw}, {Siskind}, {Spandre}, {Spinelli}, {Strong},
  {Suson}, {Tajima}, {Takahashi}, {Takahashi}, {Tanaka}, {Thayer}, {Thayer},
  {Thompson}, {Tibaldo}, {Tinivella}, {Torres}, {Tosti}, {Troja}, {Uchiyama},
  {Vandenbroucke}, {Vasileiou}, {Vianello}, {Vitale}, {Waite}, {Wallace},
  {Wang}, {Winer}, {Wood}, {Wood}, \& {Zimmer}}]{ackermann11-2lac}
{Ackermann} M. {et~al.}, 2011, \apj, 743, 171

\bibitem[{{Ackermann} {et~al}\mbox{.}(2012){Ackermann}, {Ajello}, {Allafort},
  {Antolini}, {Baldini}, {Ballet}, {Barbiellini}, {Bastieri}, {Bellazzini},
  {Berenji}, {Blandford}, {Bloom}, {Bonamente}, {Borgland}, {Bouvier},
  {Brandt}, {Bregeon}, {Brigida}, {Bruel}, {Buehler}, {Burnett}, {Buson},
  {Caliandro}, {Cameron}, {Caraveo}, {Casandjian}, {Cavazzuti}, {Cecchi}, {{\c
  C}elik}, {Charles}, {Chekhtman}, {Chen}, {Cheung}, {Chiang}, {Ciprini},
  {Claus}, {Cohen-Tanugi}, {Conrad}, {Cutini}, {de Angelis}, {DeCesar}, {De
  Luca}, {de Palma}, {Dermer}, {Silva}, {Drell}, {Drlica-Wagner}, {Dubois},
  {Enoto}, {Favuzzi}, {Fegan}, {Ferrara}, {Focke}, {Fortin}, {Fukazawa},
  {Funk}, {Fusco}, {Gargano}, {Gasparrini}, {Gehrels}, {Germani}, {Giglietto},
  {Giordano}, {Giroletti}, {Glanzman}, {Godfrey}, {Grenier}, {Grondin},
  {Grove}, {Guillemot}, {Guiriec}, {Gustafsson}, {Hadasch}, {Hanabata},
  {Harding}, {Hayashida}, {Hays}, {Healey}, {Hill}, {Horan}, {Hou},
  {J{'o}hannesson}, {Johnson}, {Johnson}, {Kamae}, {Katagiri}, {Kataoka},
  {Kerr}, {Kn{\o}dlseder}, {Kuss}, {Lande}, {Latronico}, {Lee},
  {Lemoine-Goumard}, {Longo}, {Loparco}, {Lott}, {Lovellette}, {Lubrano},
  {Madejski}, {Mazziotta}, {McEnery}, {Mehault}, {Michelson}, {Mignani},
  {Mitthumsiri}, {Mizuno}, {Monte}, {Monzani}, {Morselli}, {Moskalenko},
  {Murgia}, {Nakamori}, {Naumann-Godo}, {Nolan}, {Norris}, {Nuss}, {Ohsugi},
  {Okumura}, {Omodei}, {Orlando}, {Ormes}, {Ozaki}, {Paneque}, {Panetta},
  {Parent}, {Pelassa}, {Pesce-Rollins}, {Pierbattista}, {Piron}, {Pivato},
  {Porter}, {Rain{`o}}, {Rando}, {Ray}, {Razzano}, {Reimer}, {Reimer},
  {Reposeur}, {Romani}, {Sadrozinski}, {Salvetti}, {Saz Parkinson}, {Schalk},
  {Sgr{`o}}, {Shaw}, {Siskind}, {Smith}, {Spandre}, {Spinelli}, {Suson},
  {Takahashi}, {Tanaka}, {Thayer}, {Thayer}, {Thompson}, {Tibaldo}, {Tibolla},
  {Torres}, {Tosti}, {Tramacere}, {Troja}, {Usher}, {Vandenbroucke},
  {Vasileiou}, {Vianello}, {Vilchez}, {Vitale}, {Waite}, {Wallace}, {Wang},
  {Winer}, {Wolff}, {Wood}, {Wood}, {Yang}, \& {Zimmer}}]{ackermann12unassoc}
{Ackermann} M. {et~al.}, 2012, \apj, 753, 83




\bibitem[{{Blanchard} {et~al}\mbox{.}(2012){Blanchard}, {Lovell}, {Ojha},
  {Kadler}, {Dickey}, \& {Edwards}}]{blanchard12}
{Blanchard} J.~M., {Lovell} J.~E.~J., {Ojha} R., {Kadler} M., {Dickey} J.~M.,
  {Edwards} P.~G., 2012, \aap, 538, A150

\bibitem[{{Bloom} {et~al}\mbox{.}(2002){Bloom}, {Kulkarni}, \&
  {Djorgovski}}]{bloom02}
{Bloom} J.~S., {Kulkarni} S.~R., {Djorgovski} S.~G., 2002, \aj, 123, 1111

\bibitem[{{Cheung} {et~al}\mbox{.}(2012){Cheung}, {Donato}, {Gehrels},
  {Sokolovsky}, \& {Giroletti}}]{cheung12}
{Cheung} C.~C., {Donato} D., {Gehrels} N., {Sokolovsky} K.~V., {Giroletti} M.,
  2012, \apj, 756, 33
  
\bibitem[{{D'Ammando} {et~al}\mbox{.}(2012){D'Ammando}, {Orienti}, {Finke},
  {Raiteri}, {Angelakis}, {Fuhrmann}, {Giroletti}, {Hovatta}, {Max-Moerbeck},
  {Perkins}, {Readhead}, {Richards}, {Stawarz}, \& {Donato}}]{dammando12}
{D'Ammando} F. {et~al.}, 2012, \mnras, 426, 317

\bibitem[{{Dormody} {et~al}\mbox{.}(2011){Dormody}, {Johnson}, {Atwood},
  {Belfiore}, {Grenier}, {Johnson}, {Razzano}, \& {Saz Parkinson}}]{dormody11}
{Dormody} M., {Johnson} R.~P., {Atwood} W.~B., {Belfiore} A., {Grenier} I.~A.,
  {Johnson} T.~J., {Razzano} M., {Saz Parkinson} P.~M., 2011, \apj, 742, 126

\bibitem[{{Evans} {et~al}\mbox{.}(2009){Evans}, {Beardmore}, {Page}, {Osborne},
  {O'Brien}, {Willingale}, {Starling}, {Burrows}, {Godet}, {Vetere}, {Racusin},
  {Goad}, {Wiersema}, {Angelini}, {Capalbi}, {Chincarini}, {Gehrels}, {Kennea},
  {Margutti}, {Morris}, {Mountford}, {Pagani}, {Perri}, {Romano}, \&
  {Tanvir}}]{evans09}
{Evans} P.~A. {et~al.}, 2009, \mnras, 397, 1177

\bibitem[{{Finke}(2013)}]{finke13}
{Finke} J.~D., 2013, \apj, 763, 134

\bibitem[{{Foschini} {et~al}\mbox{.}(2011){Foschini}, {Ghisellini}, {Kovalev},
  {Lister}, {D'Ammando}, {Thompson}, {Tramacere}, {Angelakis}, {Donato},
  {Falcone}, {Fuhrmann}, {Hauser}, {Kovalev}, {Mannheim}, {Maraschi},
  {Max-Moerbeck}, {Nestoras}, {Pavlidou}, {Pearson}, {Pushkarev}, {Readhead},
  {Richards}, {Stevenson}, {Tagliaferri}, {Tibolla}, {Tavecchio}, \&
  {Wagner}}]{foschini11}
{Foschini} L. {et~al.}, 2011, \mnras, 413, 1671

\bibitem[{{Fossati} {et~al}\mbox{.}(1998){Fossati}, {Maraschi}, {Celotti},
  {Comastri}, \& {Ghisellini}}]{fossati98}
{Fossati} G., {Maraschi} L., {Celotti} A., {Comastri} A., {Ghisellini} G.,
  1998, \mnras, 299, 433

\bibitem[{{Ghisellini} \& {Tavecchio}(2008)}]{ghisellini08}
{Ghisellini} G., {Tavecchio} F., 2008, \mnras, 387, 1669

\bibitem[{{Giommi} {et~al}\mbox{.}(2012){Giommi}, {Polenta},
  {L{\"a}hteenm{\"a}ki}, {Thompson}, {Capalbi}, {Cutini}, {Gasparrini},
  {Gonz{\'a}lez-Nuevo}, {Le{\'o}n-Tavares}, {L{\'o}pez-Caniego}, {Mazziotta},
  {Monte}, {Perri}, {Rain{\`o}}, {Tosti}, {Tramacere}, {Verrecchia}, {Aller},
  {Aller}, {Angelakis}, {Bastieri}, {Berdyugin}, {Bonaldi}, {Bonavera},
  {Burigana}, {Burrows}, {Buson}, {Cavazzuti}, {Chincarini}, {Colafrancesco},
  {Costamante}, {Cuttaia}, {D'Ammando}, {de Zotti}, {Frailis}, {Fuhrmann},
  {Galeotta}, {Gargano}, {Gehrels}, {Giglietto}, {Giordano}, {Giroletti},
  {Keih{\"a}nen}, {King}, {Krichbaum}, {Lasenby}, {Lavonen}, {Lawrence},
  {Leto}, {Lindfors}, {Mandolesi}, {Massardi}, {Max-Moerbeck}, {Michelson},
  {Mingaliev}, {Natoli}, {Nestoras}, {Nieppola}, {Nilsson}, {Partridge},
  {Pavlidou}, {Pearson}, {Procopio}, {Rachen}, {Readhead}, {Reeves}, {Reimer},
  {Reinthal}, {Ricciardi}, {Richards}, {Riquelme}, {Saarinen}, {Sajina},
  {Sandri}, {Savolainen}, {Sievers}, {Sillanp{\"a}{\"a}}, {Sotnikova},
  {Stevenson}, {Tagliaferri}, {Takalo}, {Tammi}, {Tavagnacco}, {Terenzi},
  {Toffolatti}, {Tornikoski}, {Trigilio}, {Turunen}, {Umana}, {Ungerechts},
  {Villa}, {Wu}, {Zacchei}, {Zensus}, \& {Zhou}}]{giommi12}
{Giommi} P. {et~al.}, 2012, \aap, 541, A160

\bibitem[{{Grupe} {et~al}\mbox{.}(2004){Grupe}, {Leighly}, {Burwitz},
  {Predehl}, \& {Mathur}}]{grupe04}
{Grupe} D., {Leighly} K.~M., {Burwitz} V., {Predehl} P., {Mathur} S., 2004,
  \aj, 128, 1524

\bibitem[{{Kalberla} {et~al}\mbox{.}(2005){Kalberla}, {Burton}, {Hartmann},
  {Arnal}, {Bajaja}, {Morras}, \& {P{\"o}ppel}}]{kalberla05}
{Kalberla} P.~M.~W., {Burton} W.~B., {Hartmann} D., {Arnal} E.~M., {Bajaja} E.,
  {Morras} R., {P{\"o}ppel} W.~G.~L., 2005, \aap, 440, 775

\bibitem[{{Maraschi} \& {Haardt}(1997)}]{maraschi97}
{Maraschi} L., {Haardt} F., 1997, in Astronomical Society of the Pacific
  Conference Series, Vol. 121, IAU Colloq. 163: Accretion Phenomena and Related
  Outflows, {Wickramasinghe} D.~T., {Bicknell} G.~V., {Ferrario} L., eds., p.
  101

\bibitem[{{Massaro} {et~al}\mbox{.}(2012{\natexlab{a}}){Massaro}, {D'Abrusco},
  {Tosti}, {Ajello}, {Gasparrini}, {Grindlay}, \& {Smith}}]{massaro12a}
{Massaro} F., {D'Abrusco} R., {Tosti} G., {Ajello} M., {Gasparrini} D.,
  {Grindlay} J.~E., {Smith} H.~A., 2012{\natexlab{a}}, \apj, 750, 138

\bibitem[{{Massaro} {et~al}\mbox{.}(2012{\natexlab{b}}){Massaro}, {D'Abrusco},
  {Tosti}, {Ajello}, {Paggi}, \& {Gasparrini}}]{massaro12b}
{Massaro} F., {D'Abrusco} R., {Tosti} G., {Ajello} M., {Paggi} A., {Gasparrini}
  D., 2012{\natexlab{b}}, \apj, 752, 61

\bibitem[{{Massaro} {et~al}\mbox{.}(2013){Massaro}, {D'Abrusco}, {Paggi},
  {Masetti}, {Giroletti}, {Tosti}, {Smith}, \& {Funk}}]{massaro13}
{Massaro} F., {D'Abrusco} R., {Paggi} A., {Masetti} N., {Giroletti} M., {Tosti}
  G., {Smith} H.~A., {Funk} S., 2013, \apjs, 206, 13

\bibitem[{{Mirabal} {et~al}\mbox{.}(2012){Mirabal}, {Fr{\'{\i}}as-Martinez},
  {Hassan}, \& {Fr{\'{\i}}as-Martinez}}]{mirabal12}
{Mirabal} N., {Fr{\'{\i}}as-Martinez} V., {Hassan} T., {Fr{\'{\i}}as-Martinez}
  E., 2012, \mnras, 424, L64

\bibitem[{{Nolan} {et~al}\mbox{.}(2012){Nolan}, {Abdo}, {Ackermann}, {Ajello},
  {Allafort}, {Antolini}, {Atwood}, {Axelsson}, {Baldini}, {Ballet}, \&
  et~al.}]{nolan12}
{Nolan} P.~L. {et~al.}, 2012, \apjs, 199, 31

\bibitem[{{Ojha} {et~al}\mbox{.}(2010){Ojha}, {Kadler}, {B{\"o}ck}, {Booth},
  {Dutka}, {Edwards}, {Fey}, {Fuhrmann}, {Gaume}, {Hase}, {Horiuchi},
  {Jauncey}, {Johnston}, {Katz}, {Lister}, {Lovell}, {M{\"u}ller}, {Pl{\"o}tz},
  {Quick}, {Ros}, {Taylor}, {Thompson}, {Tingay}, {Tosti}, {Tzioumis}, {Wilms},
  \& {Zensus}}]{Ojha10}
{Ojha} R. {et~al.}, 2010, \aap, 519, A45

\bibitem[{{Petrov} {et~al}\mbox{.}(2013){Petrov}, {Mahony}, {Edwards},
  {Sadler}, {Schinzel}, \& {McConnell}}]{petrov13}
{Petrov} L., {Mahony} E.~K., {Edwards} P.~G., {Sadler} E.~M., {Schinzel} F.~K.,
  {McConnell} D., 2013, \mnras, 432, 1294

\bibitem[{{Ray} {et~al}\mbox{.}(2012){Ray}, {Abdo}, {Parent}, {Bhattacharya},
  {Bhattacharyya}, {Camilo}, {Cognard}, {Theureau}, {Ferrara}, {Harding},
  {Thompson}, {Freire}, {Guillemot}, {Gupta}, {Roy}, {Hessels}, {Johnston},
  {Keith}, {Shannon}, {Kerr}, {Michelson}, {Romani}, {Kramer}, {McLaughlin},
  {Ransom}, {Roberts}, {Saz Parkinson}, {Ziegler}, {Smith}, {Stappers},
  {Weltevrede}, \& {Wood}}]{ray12}
{Ray} P.~S. {et~al.}, 2012, ArXiv e-prints

\bibitem[{{Romani}(2012)}]{romani12}
{Romani} R.~W., 2012, \apjl, 754, L25

\bibitem[{{Stevens} {et~al}\mbox{.}(2012){Stevens}, {Edwards}, {Ojha},
  {Kadler}, {Hungwe}, {Dutka}, {Tingay}, {Macquart}, {Moin}, {Lovell}, \&
  {Blanchard}}]{stevens12}
{Stevens} J. {et~al.}, 2012, arXiv: 1205.2403

\bibitem[{{Stratta} {et~al}\mbox{.}(2011){Stratta}, {et al.}, \& {on behalf of
  the ASDC team}}]{stratta11}
{Stratta} G., {et al.}, {on behalf of the ASDC team}, 2011, arXiv:1103.0749


\bibitem[{{Stroh} \& {Falcone}(2013)}]{stroh13}
{Stroh} M.C., {Falcone} A.D., 2013, \apjs, 207, 28


\bibitem[{{Su} \& {Finkbeiner}(2012)}]{su12}
{Su} M., {Finkbeiner} D.~P., 2012, arXiv:1207.7060

\bibitem[{{Tornikoski} {et~al}\mbox{.}(2002){Tornikoski},
  {L{\"a}hteenm{\"a}ki}, {Lainela}, \& {Valtaoja}}]{tornikoski02}
{Tornikoski} M., {L{\"a}hteenm{\"a}ki} A., {Lainela} M., {Valtaoja} E., 2002,
  \apj, 579, 136

\end{thebibliography}

\end{document}